\documentclass[aps,prb,twocolumn,superscriptaddress,10pt]{revtex4-1}

\usepackage{graphicx}
\usepackage{amsmath}
\usepackage{amssymb}
\usepackage{amsfonts}
\usepackage{bm}
\usepackage{psfrag}
\usepackage{pstricks}

\usepackage{color}

\pdfoutput=1

\begin{document}

\title{Random alloy fluctuations and structural inhomogeneities in
$c$-plane In$_{x}$Ga$_{1-x}$N quantum wells: theory of ground and excited electron and hole states.}

\author{Daniel S.~P.~Tanner}
\affiliation{Photonics Theory Group, Tyndall National Institute,
Dyke Parade, Cork, Ireland} \affiliation{Department of Physics,
University College Cork, Cork, Ireland}

\author{Miguel~A. Caro}
 \affiliation{Department of
 Electrical Engineering and Automation, Aalto University, Espoo,
 Finland} \affiliation{COMP Centre of Excellence in Computational
 Nanoscience, Aalto University, Espoo, Finland}

\author{Eoin~P. O'Reilly}
\affiliation{Photonics Theory Group, Tyndall National Institute,
Dyke Parade, Cork, Ireland} \affiliation{Department of Physics,
University College Cork, Cork, Ireland} 
 
\author{Stefan Schulz}
\affiliation{Photonics Theory Group, Tyndall National Institute,
Dyke Parade, Cork, Ireland} 

\begin{abstract}
We present a detailed theoretical analysis of the electronic
structure of $c$-plane InGaN/GaN quantum wells with indium contents
varying between 10\% and 25\%. The electronic structure of the
quantum wells is treated by means of an atomistic tight-binding
model, accounting for variations in strain and built-in field due to
random alloy fluctuations. Our analysis reveals strong localisation
effects in the hole states. These effects are found not only in the
ground states, but also the excited states. We conclude that
localisation effects persist to of order 100~meV into the valence
band, for as little as 10\% indium in the quantum well, giving rise
to a significant density of localised states. We find, from an
examination of the modulus overlap of the wave functions, that the
hole states can be divided into three regimes of localisation. Our
results also show that localisation effects due to random alloy
fluctuations are far less pronounced for electron states. However,
the combination of electrostatic built-in field, alloy fluctuations
and structural inhomogeneities, such as well-width fluctuations, can
nevertheless lead to significant localisation effects in the
electron states, especially at higher indium contents. Overall, our
results are indicative of individually localised electron and hole
states, consistent with the experimentally proposed explanation of
time-dependent
photoluminescence results in $c$-plane InGaN/GaN QWs.
\end{abstract}

\date{\today}



\maketitle

\section{Introduction}

Nitride semiconductors have attracted considerable interest for a variety of different
applications, ranging from photovoltaic cells up to optoelectronic devices such as
light-emitting devices (LEDs).~\cite{Za2016}
For instance, the cornerstone of modern LEDs operating in
the blue to green spectral region are InGaN/GaN quantum wells (QWs) grown
along the crystallographic $c$ axis.~\cite{NaCh00,PoBo1997,LiJi15,Humphreys_MRS} The success of these devices is
remarkable given the extremely high defect densities ($>10^8$
cm$^{-2}$) in InGaN/GaN materials, which mainly originate from the
large lattice mismatch between GaN and the underlying substrate,
sapphire (16\%).~\cite{ChUe06,Humph_RSC} The widely accepted
explanation for the defect-insensitive efficiency of InGaN - based
devices is that carrier localisation effects introduced by alloy fluctuations
prevent diffusion to non-radiative recombination centres. Using positron
annihilation measurements,
Chichibu \emph{et al.}~\cite{ChUe06} demonstrated the presence of such carrier
localisation effects in nitride-based alloys. In addition to these
measurements, photoluminescence (PL) spectroscopy studies by
different groups also gave clear indications that the optical
properties of $c$-plane InGaN/GaN QWs are significantly affected by
localisation phenomena. For example, temperature dependent PL measurements
have shown that the PL peak energy follows an ``S-shaped'' temperature
dependence.~\cite{HaWa12,ChGa98,ElPe97,LiJi15} This particular form of the shift of the PL peak position is attributed
to the redistribution of carriers between different localised
states.~\cite{ElPe97,PeOs96,LiXu01} Furthermore, time-dependent PL measurements revealed
non-exponential decay transients and that the decay times extracted
from the curves vary across the PL curve.~\cite{DaDa00,DaSc16,ChGa98} Morel~\emph{et al.}~\cite{MoLe02}
proposed as an explanation for this that the radiative recombination
process in $c$-plane InGaN/GaN QWs is dominated by individually localised
carriers. The varying spatial separations of these separately localised
carriers, both in the $c$-plane and perpendicular to it,
lead to variations in the radiative recombination time. Using this
assumption, Morel~\emph{et al.}~\cite{MoLe02} were able to obtain a good agreement
between theoretical predictions and the experimental data.

Even though, as discussed above, there is considerable experimental
evidence for the importance and presence of localisation effects due
to alloy fluctuations, it is only recently that these effects have been
considered in theoretical studies. The applied theoretical
frameworks range from modified continuum-based descriptions~\cite{WaGo2011,YaSh2014} up to
fully atomistic models.~\cite{ScCa2015,LiLu2010,MaPe16} These studies have focused mainly on ground state
properties, which are important to understand and explain experimental studies
at low temperature; however, in order to understand the results of experiments conducted
at ambient temperature, as well as the transport properties of the system, many excited states
must be considered. Already the ``S-shaped'' temperature dependence of the PL
peak position indicates that excited states exhibit localisation
features. These localised states modify the form of the density of states
in such a way that there is a smooth tail of states at the low energy end of
the density of states in a QW structure.~\cite{Ur1953,WaCh11} Because of this, these states are
often referred to as ``tail states''.~\cite{BoVo12}

In this paper we address the impact of random alloy fluctuations on
the localisation features of both ground and excited states in
$c$-plane InGaN/GaN QW systems. To cover the experimentally
relevant indium composition ranges, we analyse InGaN/GaN QWs with
indium contents of 10\%, 15\% and 25\%. Furthermore, since
experimental studies highlight also that structural inhomogeneities,
such as well width fluctuations, impact the electronic and optical
properties of nitride-based heterostructures
significantly,\cite{SaHe14} we include these effects also in our
atomistic analysis. Our theoretical framework is based on an
atomistic $sp^3$ tight-binding (TB) model, which includes effects
such as strain and polarisation field variations due to the
considered random alloy fluctuations.

Our calculations reveal that random alloy fluctuations lead to
strong hole wave function localisation effects in both ground and
excited states. We find here that over an energy range of order $100$~meV
a significant density of localised valence states is
expected. The presented data also indicates that these localised
states significantly affect the probability for transferring
carriers from one site/state to another. When studying whether or
not the hole wave functions overlap with each other, three different
regimes become apparent. The first corresponds to
``strongly localised states'' with almost no spatial overlap with
all other states   considered. The second and third regimes consist of what we refer to as ``semi-localised states''
and ``delocalised states'', respectively, where the spatial overlap is significantly increased with respect to the
``strongly localised'' states. Our data also shows that the number of
states, and therefore the energy range, constituting each of these
regimes, depends on the indium content of the system in question.

While our calculations reveal that random alloy fluctuations lead to
very strong hole wave function localisation effects, the situation
is different for the electron states. Compared with the hole states,
the alloy fluctuations lead to much less pronounced electron wave
function perturbations. The primary sources of the localisation of
electron states are the electrostatic built-in field and well width
fluctuations present in $c$-plane InGaN/GaN heterostructures. Also
for the excited electron states, localisation effects are strongly
reduced compared with the holes.

The combination of macroscopic built-in field, random alloy and well
width fluctuations leads to a spatial separation of electron and
hole wave functions both in the $c$ plane and perpendicular to it.
In standard continuum-based models, InGaN/GaN QWs are treated as
ideal one-dimensional systems, which can be described by averaged
material parameters. These approaches can account only for the
spatial separation of electron and hole wave functions along the
growth direction due to the presence of the built-in field. Thus, in
contrast to the here applied fully atomistic three-dimensional
approach, in-plane spatial separations are not captured. Our results
indicate that electrons and holes are individually localised and
that the wave function overlap should therefore also depend on the
relative in-plane position of the carriers. Thus, the here
obtained findings are consistent with the ``pseudo 2-D
donor-acceptor pair system'' proposed by Morel \emph{et
al.}~\cite{MoLe02} to explain time-dependent PL results of $c$-plane
InGaN/GaN QWs.

The manuscript is organised as follows. In Sec.~\ref{sec:theory}, we
introduce the components of our theoretical framework. In
Sec.~\ref{sec:QWModel} we discuss the QW model system under
consideration and the input from available experimental structural
data. The results of our calculations are presented in
Sec.~\ref{sec:Results}. We first address ground state properties in
Sec.~\ref{sec:GS_prop} before turning to the excited states in
Sec.~\ref{sec:ES_prop}. We relate our theoretical data to
experimental findings in Sec.~\ref{sec:Results_Experiment}, before
summarising our work in Sec.~\ref{sec:Summary}.

\section{Theoretical Framework}
\label{sec:theory}

In this section we briefly introduce the atomistic theoretical
framework used to study the electronic structure of $c$-plane
InGaN/GaN QWs with varying indium content. Our approach can be
divided into three main components. First, the large lattice
mismatch between InN and GaN (approx. 11\%) gives rise to a strain
field in InGaN/GaN heterostructures. To treat this strain field on
an atomistic level, and thus account for random alloy fluctuations
on a microscopic level, we employ a valence-force-field (VFF) model
based on that introduced by Martin.~\cite{Ma1970} Our VFF includes
electrostatic effects explicitly and reproduces important
real-wurtzite system attributes such as non-ideal $c/a$ ratios and
internal parameters $u$. We have implemented this model in the
software package LAMMPS.~\cite{LAMMPS} More details are given in
Ref.~\citenum{ScCa2015}.

Second, the strong intrinsic electrostatic built-in
fields in nitride heterostructures have to be included to achieve a
realistic description of the electronic structure of $c$-plane
InGaN/GaN QWs. In wurtzite III-N materials the lack
of inversion symmetry leads to a non-vanishing sum of electric
dipole moments and thus to a macroscopic electric polarisation. This
polarisation has two contributions, one of which is strain
\textit{independent}, known as the spontaneous polarisation, and the
other of which is the strain \textit{dependent} piezoelectric
polarisation.~\cite{BeFi97} In addition to the macroscopic polarisation,
random alloy fluctuations lead also to local polarisation
variations. Recently, we have developed a local polarisation theory
capable of accounting for both the  macroscopic and local
intrinsic polarisation.~\cite{CaSc2013} 
Our theory receives input for its material
parameters from density functional theory (DFT) within the
Heyd-Scusera-Ernzerhof (HSE)~\cite{HeSc03} screened exchange hybrid
functional scheme. The starting point for
this approach is to split the wurtzite polarisation vector, made up
of spontaneous and piezoelectric contributions, into macroscopic and
microscopic terms. The macroscopic term is the so called clamped-ion
contribution where ions are not allowed to move and this part is
related to the piezoelectric coefficients $e^{0}_{ij}$.~\cite{BeFi97} The local contribution
involves the deformation of the nearest neighbor environment around
the atom under consideration. With this we can define a dipole
moment for each tetrahedron over the entire cell. From this we can
calculate the corresponding polarisation (dipole moment/Volume). The
last step is now to calculate the resulting built-in
potential. Usually this is done by solving Poisson's equation.
However, since we are dealing with an atomic grid this becomes
difficult. To circumvent this problem we make use of the multipole
expansion of a distribution of electric charges, with which we can
calculate the electrostatic potential at position $\mathbf{r}$ due
to the presence of a point dipole at $\mathbf{r'}$. In doing so we avoid numerical
problems arising from solving Possion's equation by, for instance, a
finite difference method on an irregular wurtzite crystal. Consequently, we
find the situation that the macroscopic component of the built-in
potential, which is effectively the potential one would expect in a
capacitor, is modified in the QW by local fluctuations
sitting on top of the potential slope in this region. The details
of the local polarisation and point dipole method are described in
Ref.~\citenum{CaSc2013} in more detail. Note that we are here interested only
in the intrinsic properties of InGaN/GaN QWs, thus we do not include
the effects of any externally applied bias, as one would have
in an LED structure.




Thirdly, to determine the effects of alloy, strain and built-in potential
fluctuations on the electronic structure of InGaN/GaN QWs, we use an
atomistic, nearest neighbor $sp^{3}$ TB model.
Before treating the InGaN alloy we start from the binary materials
InN and GaN. The required TB parameters are determined by fitting
the TB bulk band structures of InN and GaN to those calculated using
HSE hybrid-functional DFT. 
Due to the minimal basis used in $sp^{3}$ TB, the description of the conduction band at the L- and M-valleys is less accurate.
However, from our HSE-DFT calculations we find that, for the nitrides, there is a very large energetic speparation between
the conduction band minimum at the $\Gamma$-point and the $M$ and $L$ valleys. Because of this, the evolution of the energy gap is 
dominated by the band structure around $\bf{k}=0$; thus because the TB model used here captures well the valence band and the conduction band at $\Gamma$, 
it is particularly suitable for treating nitride semiconductors.

Equipped with this knowledge about the
binary materials, we can then treat the InGaN alloy on a microscopic
level. To this end, at each atomic site, the TB parameters are set
according to the bulk values of their constituent atoms.
For the cation sites (Ga,In), there is no ambiguity in assigning the
on-site and nearest neighbor TB matrix elements, since these always
have nitrogen atoms as their nearest neighbors. However, the nearest
neighbor environment of the anions (N) will vary depending on the
local indium distribution. To treat this effect,
different approaches have been used in the literature. One ansatz is
to start already at the bulk band structure level and use the same
on-site TB matrix elements for N in InN and GaN. In doing so the
ambiguity for the N-atom on-site energies in an InGaN alloy is
removed. However, by
assuming the same N-atom on-site energies in InN and GaN, one
basically fixes the band offset between InN and GaN. Such an
approach limits the transferability of the TB parameters
to other systems. Given these arguments, we apply here another widely used
approach to treat the on-site energies of a common atom species in
an alloy. The assignment here is performed using weighted averages
for the on-site energies, where the weights are determined by the
number of nearest neighbor In or Ga atoms. This is a widely used
and benchmarked approach to treat alloys in an
atomistic TB framework.~\cite{ReLi02,BoKh07,ScCa2015,Zi2013}
The band offset is included by shifting the InN
on-site TB parameters by the valence band offset $\Delta
E_\text{VB}=0.62$ eV.  The value for $\Delta E_\text{VB}$ is taken
from HSE-DFT calculations.~\cite{MoMi11}
Strain and built-in potential effects are included in the
description as on-site corrections to the TB Hamiltonian. The
procedure is detailed in Ref.~\citenum{CaSc2013}.
This framework has already been
successfully applied to other wurtzite III-N alloys, such as AlGaN
and AlInN.~\cite{ScCaPaMa13,CoSc15} And similar approaches have been used also
to effectively describe other alloy systems, such as GaBiAs/GaAs QWs.~\cite{UsRe14}

Having discussed the theoretical framework, we introduce the model
$c$-plane InGaN/GaN QW systems in the next section.

\section{InGaN QW System}
\label{sec:QWModel}

Here we introduce the QW systems to which we apply our theoretical
framework. The QW structures being studied are similar in indium
contents to QWs studied experimentally in Ref.~\citenum{GrSo05}. To
model $c$-plane In$_{x}$Ga$_{1-x}$N/GaN QWs we use $\approx$ 82,000
atom supercells (equivalent to a system size of $\approx10\,
\text{nm} \times 9\, \text{nm} \times 10\, $nm) with periodic
boundary conditions. The QW in these supercells is around 3.5 nm
wide. These supercell dimensions have been chosen such that
the experimentally reported carrier localisation lengths of
$1.1-3.1$~nm\cite{GrSo05} can be accommodated within the cell
without spurious coupling to periodic replicas. Following
Refs.~\citenum{MoPa2002,SmKa2003,GaOl2007,Humphreys2007}, 
we assume that InGaN is a random alloy and
distribute indium atoms at the cation sites of the active region
with a probability given by the nominal indium content of the
configuration in question. In doing so we do not assume any
preferential orientation or correlation of indium atoms. To examine
the impact of the microscopic indium configuration on the electronic
structure, we consider twenty different random atomic configurations
for each composition studied. These configurations are generated for
nominal indium contents of 10\% , 15\% and 25\% ,
which cover the experimentally relevant range of
indium contents.~\cite{GrSo05} It should be noted that we are interested
here in general trends rather than a detailed statistical analysis
of the results. Such an analysis would require significantly more random configurations. 
However, for our purposes, to shed light on trends and basic properties of the InGaN/GaN
QWs with varying indium content, including effects of random alloy
fluctuations, a sample of twenty different configurations per alloy
content is sufficient.

In Refs.~\citenum{GrSo05} and~\citenum{GaOl08},
well-width fluctuations were observed at the upper interface [GaN on
InGaN] of $c$-plane InGaN/GaN QWs. The reported diameters range from
5 to 10 nm and heights of one to two monolayers are observed. We
include a well-width fluctuation with a diameter of 5 nm and a
height of two monolayers sitting in the center of the upper region
of our QW. Consistent with previous approaches to the modeling of
these features, the shape of the well-width fluctuation is assumed
to be disklike.~\cite{ScCa2015,WaGo2011} We are mainly interested in
the impact of random alloy fluctuations on the electronic
properties, therefore, we do not attempt to study the effects of
well width fluctuations varying in size and shape. This would
require very detailed experimental information as input into our
model and is beyond the scope of the present work. However, assuming
a single type of well width fluctuation does give an indication of
the mechanisms by which any well-width fluctuation, in combination
with random alloy and built-in field effects, could affect carrier
localisation features.

\section{Results}
\label{sec:Results}

In this section we present the results of our theoretical analysis.
We start with ground state properties in Sec.~\ref{sec:GS_prop}, and
focus then in detail on excited states in Sec.~\ref{sec:ES_prop}.

\subsection{Ground state properties}
\label{sec:GS_prop}

\begin{figure}[t]
\centering\includegraphics[width=0.95\columnwidth]{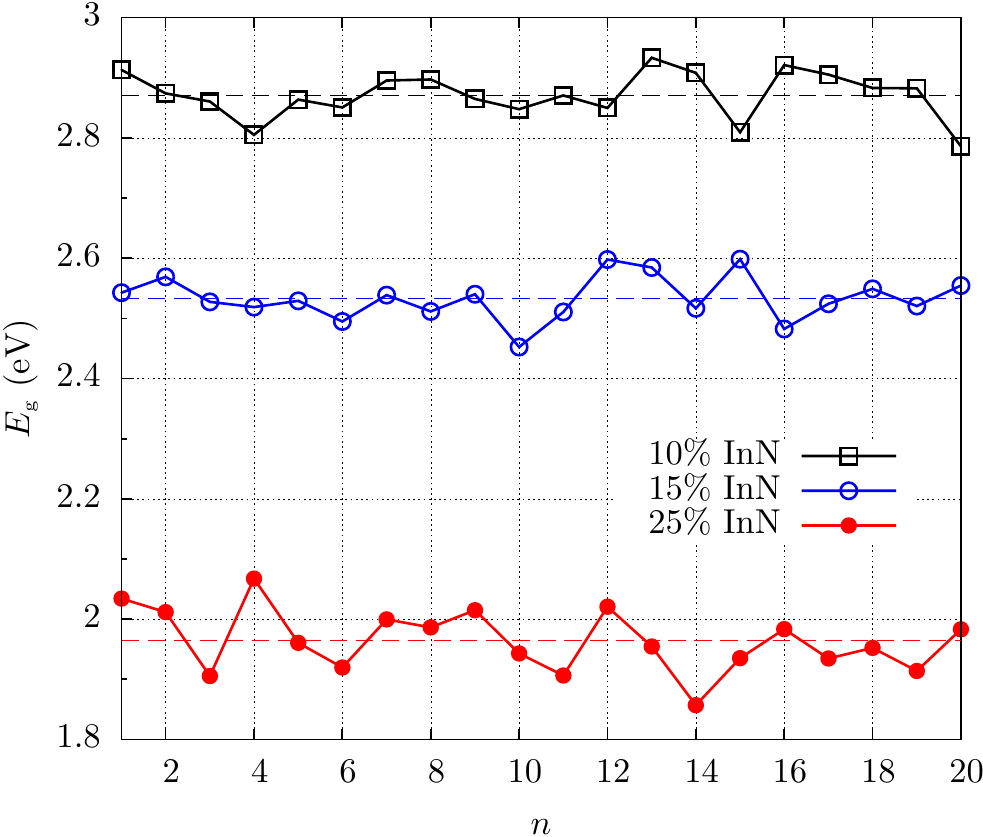}
\caption{(Color online) Single-particle ground state transition
energies in the here-considered $c$-plane In$_x$Ga$_{1-x}$N/GaN QWs
for the $n$ different random microscopic configurations.
The indium content in the well is $x=0.1$ (10\%,
black square), $x=0.15$ (15\%, open blue circle) and $x=0.25$ (25\%, red
solid circle). The average transition energies for each indium content are indicated
by the dashed lines.} \label{fig:GS_tran}
\end{figure}



A first quantitative measure for the impact of alloy fluctuations on
the electronic structure of $c$-plane InGaN/GaN QW systems is given
by the variation in ground state transition energies about their
configurational average. In Fig. \ref{fig:GS_tran}, the energy of
the ground state transition, $E_{g}$, is plotted against the
configuration number, $n$, for each nominal indium concentration
[10\%, 15\% and 25\%].
For each case the average transition energy is indicated by a dashed
line. The values obtained for the 10\% (black
square), 15\% (open blue circle) and 25\% (solid red circle) indium
systems are $2.871~$eV, $2.533~$eV and $1.964~$eV, respectively. As
expected, the transition energy shifts to lower energies with
increasing indium content.
Apparent from Fig.~\ref{fig:GS_tran} is the significant spread in
transition energies about their averages across all indium contents.
This demonstrates that random alloy effects are important for as
little as 10\% indium in the well. The fluctuations in $E_g$ are
also consistent with the large PL linewidths observed
experimentally.~\cite{DaDa00,GrSo05,ChAb98,BrWa02,DoMa99}
The here calculated average transition energies
will be compared to experimental PL peak energies reported in the
literature in Sec.~\ref{sec:Results_Experiment}.

\begin{figure}[t!]
\centering\includegraphics[width=0.95\columnwidth]{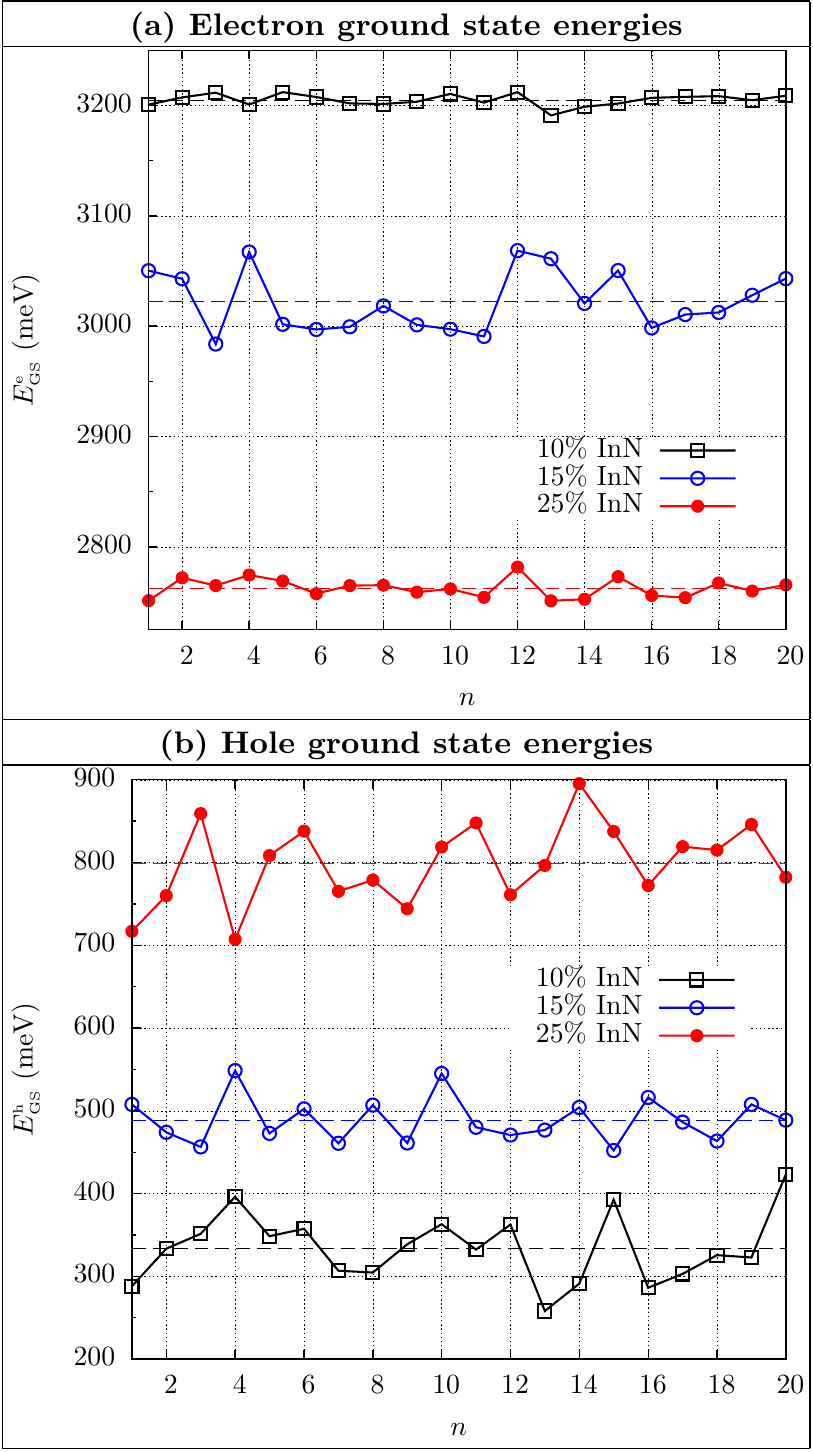}
\caption{(Color online) (a), Electron ground state
energy, $E_{\textrm{\tiny{GS}}}^{\textrm{\tiny{e}}}$, and (b), hole ground state energy, $E_{\textrm{\tiny{GS}}}^{\textrm{\tiny{h}}}$,  in an In$_{x}$Ga$_{1-x}$N/GaN QW as function of the
different random microscopic configurations, $n$. The
indium content $x$ in the well is $x=0.10$ (10\%, black square),
$x=0.15$ (15\%, blue circle) and $x=0.25$ (25\%, red solid circle). The average
ground state energies are indicated by dashed lines.}
\label{fig:electron_hole_vary}
\end{figure}

To investigate the origin of the variance of the transition
energies, we look now to the variation of the corresponding electron
and hole ground state energies about their respective averages. In
Fig.~\ref{fig:electron_hole_vary} the electron ground state
energies, $E_{\textrm{\tiny{GS}}}^{\textrm{\tiny{e}}}$
[Fig.~\ref{fig:electron_hole_vary} (a)], and hole ground state
energies, $E_{\textrm{\tiny{GS}}}^{\textrm{\tiny{h}}}$
[Fig.~\ref{fig:electron_hole_vary} (b)], are plotted as a function
of the configuration number, $n$. The average ground state energies
are again indicated as dashed lines. The data are shown for the
different indium contents $x$.
The average hole ground state energy increases with increasing
indium content since the valence band offset increases. For 10\%
[$x=0.1$, black square], 15\% [$x=0.15$, blue circle] and 25\%
[$x=0.25$, red solid circle] indium the average energies are
$334$~meV, $489$~meV and $785$~meV, respectively [cf.
Fig.~\ref{fig:electron_hole_vary} (b)]. The zero of energy is
taken as the valence band edge of the unstrained bulk GaN.
However, due to the macroscopic built-in potential and random
alloy fluctuations, the valence and conduction band
edges will vary with position across the QW structure, with the calculated hole energies then including a contribution
from these factors. For the comparison with the experiment only the transition energies are relevant, which are independent
of the choice of the zero of energy. The average ground state energies of the
electrons likewise decrease with increasing indium content, since
the conduction band offset is increased [cf.
Fig.~\ref{fig:electron_hole_vary} (a)]. The average electron ground
state energies are $3.205$~eV, $3.022$~eV and $2.763$~eV for 10\%,
15\% and 25\% indium, respectively.
\begin{figure*}[t!]
\centering\includegraphics{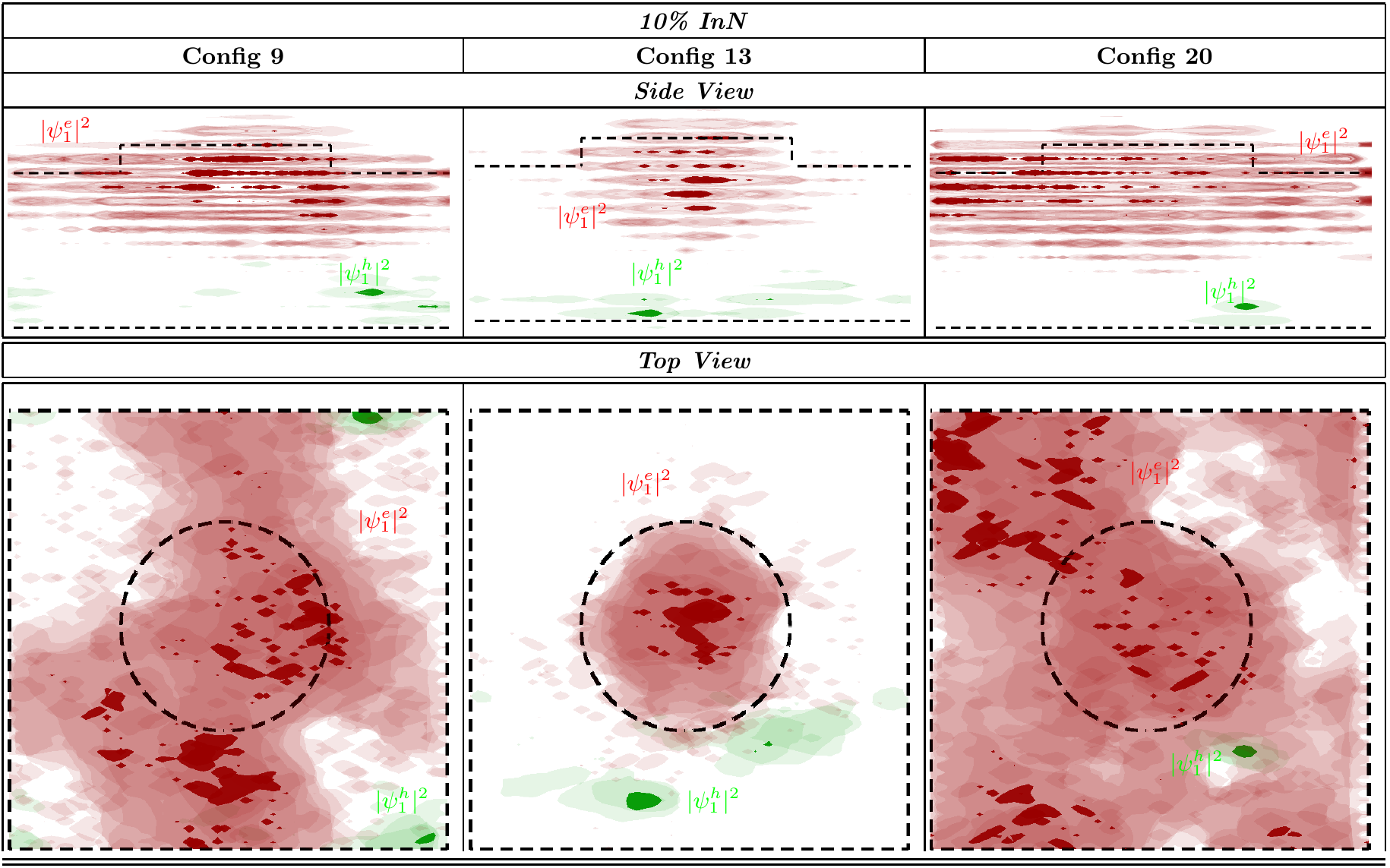}
\caption{(Color online) Isosurface plots of the electron (red), $\lvert\psi^{e}_{1}\rvert^{2}$, and
hole (green), $\lvert\psi^{e}_{1}\rvert^{2}$, ground state charge densities in the
In$_{0.10}$Ga$_{0.90}$N/GaN QW. The light (dark) isosurface
corresponds to 10\% (50\%) of the maximum charge density. The
results are shown perpendicular (``Side View'') and parallel (``Top
View'') to the $c$-axis. In this and the following two figures, the dashed lines in the side view indicate the QW
interfaces; as a guide to the eye, the circular well width
fluctuation is also given by the dashed line in the ``Top View''.  The results are
shown for three different random microscopic configurations (Configs
9, 13, and 20).}\label{fig:10_percent_WFs}
\end{figure*}
On comparison of the variations in Fig.~\ref{fig:electron_hole_vary}
(a) and (b), in general, it is evident that the hole ground state
energy, $E_{\textrm{\tiny{GS}}}^{\textrm{\tiny{h}}}$, is very
sensitive to the configuration number, $n$. This
further indicates that the alloy microstructure plays an important
role for the valence band states. More specifically, from
Fig.~\ref{fig:electron_hole_vary} we find that the
$E_{\textrm{\tiny{GS}}}^{\textrm{\tiny{h}}}$ vary between $\sim
\pm100$~meV around their average energies whilst the electron ground
state energies, $E_{\textrm{\tiny{GS}}}^{\textrm{\tiny{e}}}$, vary
at most from the average by $\sim \pm50$~meV.
However, for the electron ground states, the large value of $\pm 50$
meV arises mainly from the 15\% indium case while for 10\% and 25\%
indium we find only $\sim \pm 10$ meV and $\sim \pm 20$ meV,
respectively. Furthermore, for the 15\% indium case, the spread in
the electron ground state energies about their average energy is
comparable with the spread in the energies of the hole states.

\begin{figure*}[t!]
\centering\includegraphics{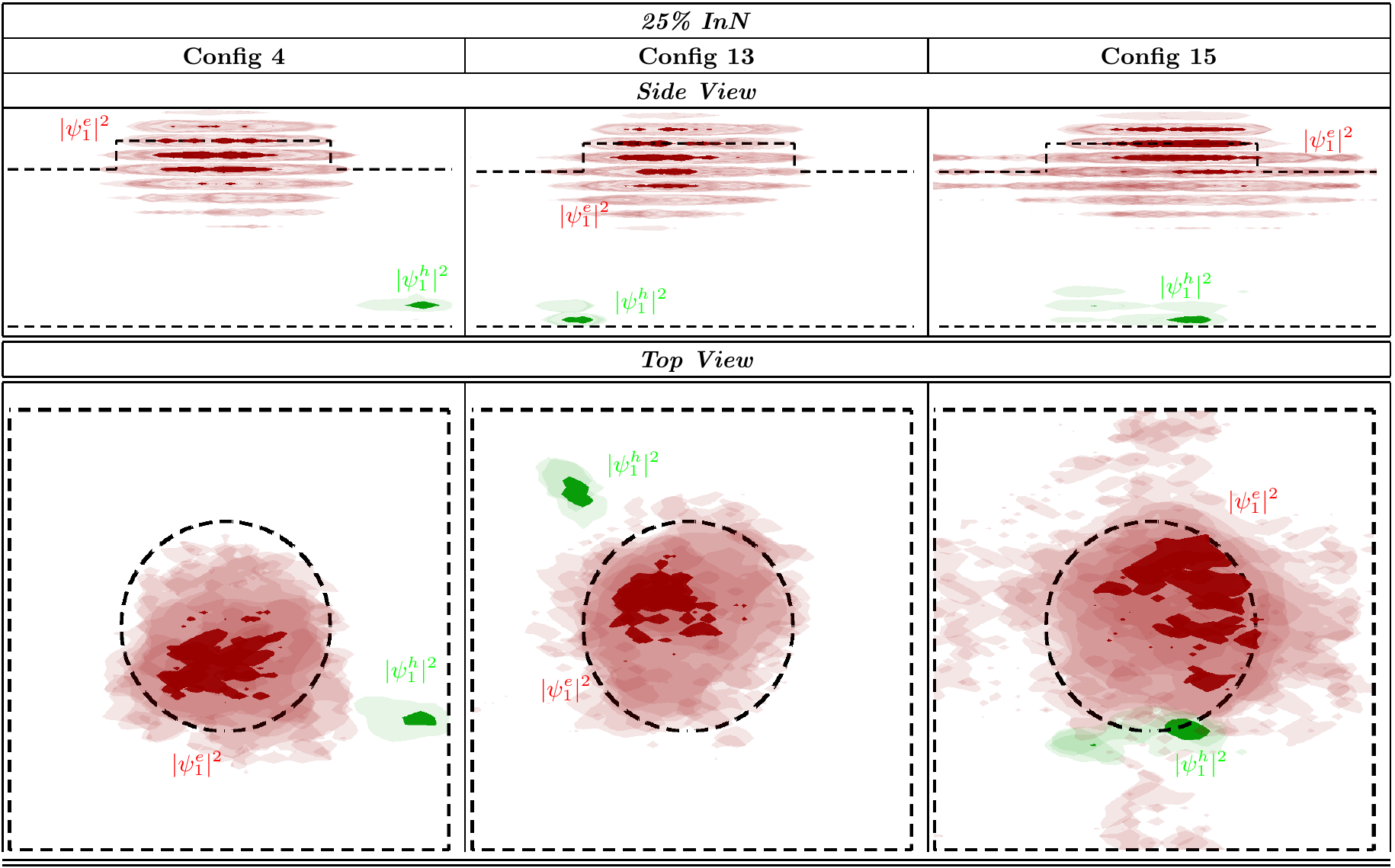}
\caption{(Color online) Isosurface plots of the electron (red), $\lvert\psi^{e}_{1}\rvert^{2}$, and
hole (green), $\lvert\psi^{e}_{1}\rvert^{2}$, ground state charge densities in the
In$_{0.25}$Ga$_{0.75}$N/GaN QW. The light (dark) isosurface
corresponds to 10\% (50\%) of the maximum charge density. The
results are shown perpendicular (``Side View'') and parallel (``Top
View'') to the $c$-axis for three different random microscopic configurations (Config
4, 13, and 15).}\label{fig:25_percent_WFs}
\end{figure*}
\begin{figure*}[t!]
\centering\includegraphics{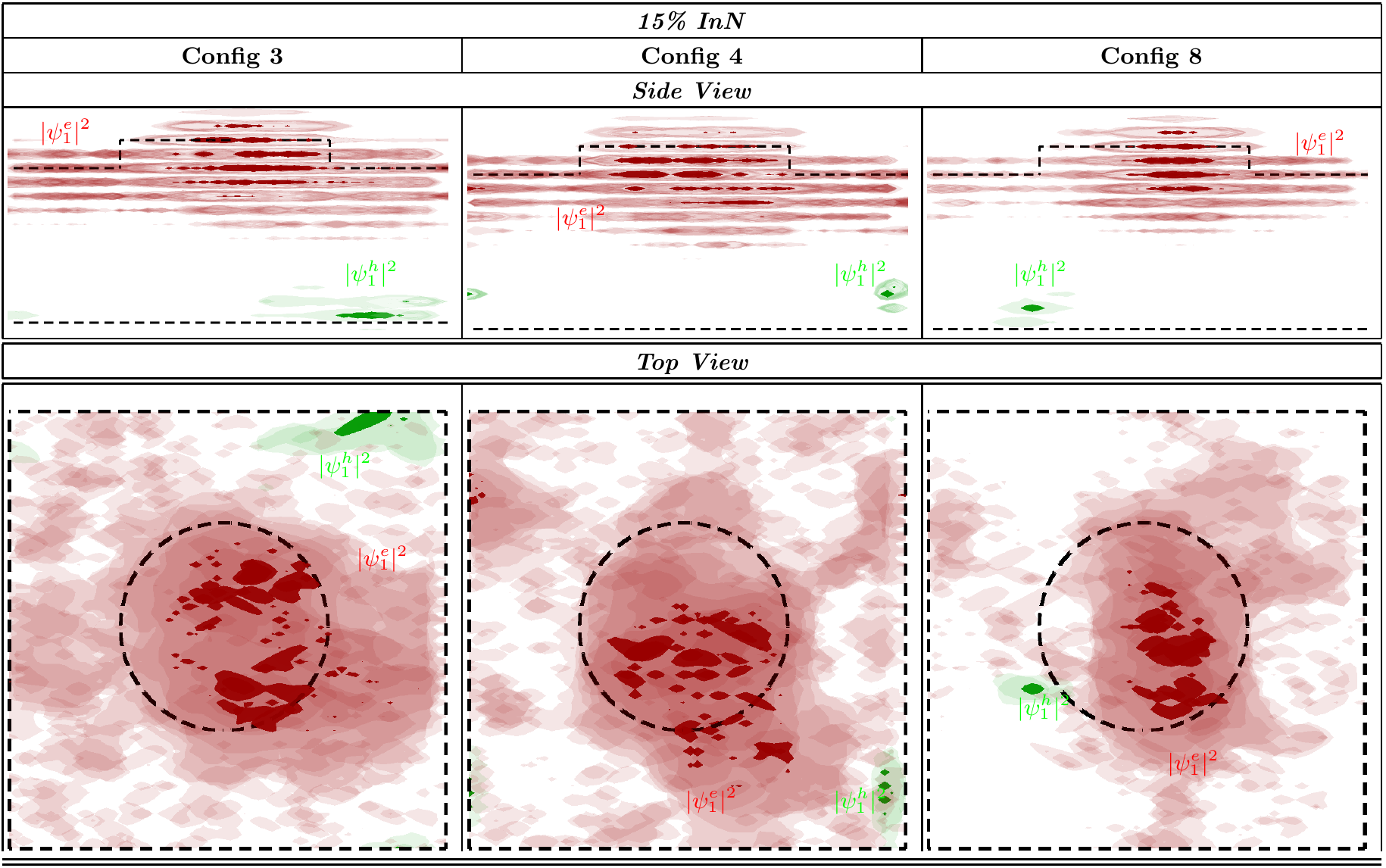}
\caption{(Color online) Isosurface plots of the electron (red), $\lvert\psi^{e}_{1}\rvert^{2}$, and
hole (green), $\lvert\psi^{e}_{1}\rvert^{2}$, ground state charge densities in the
In$_{0.15}$Ga$_{0.85}$N/GaN QW. The light (dark) isosurface
corresponds to 10\% (50\%) of the maximum charge density. The
results are shown perpendicular (``Side View'') and parallel (``Top
View'') to the $c$-axis for three different random microscopic configurations (Config
3, Config 4, Config 8).} \label{fig:15_percent_WFs}
\end{figure*}

To shed more light on the results shown in
Fig.~\ref{fig:electron_hole_vary} (a) and (b),
Figs.~\ref{fig:10_percent_WFs},~\ref{fig:25_percent_WFs}
and~\ref{fig:15_percent_WFs} show isosurfaces of the electron (red)
and hole (green) ground state charge densities for selected
configurations in the case of 10\%, 25\% and 15\% indium in the QW,
respectively. The ``Side View'' for each of these cases is a view
perpendicular to the $c$-axis, while the ``Top View'' is a view
looking down the $c$-axis. The light and dark isosurfaces correspond
to 10\% and 50\% of the respective maximum charge density values.
The selected configurations correspond to situations
with positive and negative deviations from the average ground state
energy plus one configuration that is close to the average value.

We analyse in a first step configurations $n=9, 13$ and 20 of the
10\% indium case. The corresponding charge densities are displayed
in Fig.~\ref{fig:10_percent_WFs}. 
In general we find that the electron and hole wave functions are
spatially separated along the $c$-axis due to the presence of the
electrostatic built-in field. Looking at configurations 9 and 20, we
see from the electron charge densities that they are almost spread
across the entire $c$-plane in the QW region. However, a closer
inspection also reveals that the ground state electron wave
functions are affected by the presence of the random alloy
fluctuations. This is evinced by the lower probability densities in
certain parts of the QW region. For configurations 9 and 20 we find
also that the assumed well width fluctuation is of secondary
importance. A different behaviour in our
calculations is observed when looking at configuration 13, where
the charge density is localised very strongly in the well-width
fluctuation. This particularity of the ground state wave function,
confined in the well-width fluctuation, is also reflected in the
energy value, which shows the largest absolute deviation from the
average [cf. Fig.~\ref{fig:electron_hole_vary} (a)]. However, we
stress here again that we have assumed only a particular type of
well-width fluctuation. We will discuss the importance of the
well-width fluctuation in more detail below.


For the hole ground states the situation is different. Looking at
the charge densities (green isosurfaces) in
Fig.~\ref{fig:10_percent_WFs}, much stronger localisation effects
are visible for all configurations. The ``strength'' and spatial
position of the localisation changes greatly from configuration to
configuration. This behaviour reflects the sensitivity of the hole
ground state energies to a particular microscopic configuration, as
seen in Fig.~\ref{fig:electron_hole_vary} (b). In general this
sensitivity to the alloy microstructure can be attributed to the
larger effective mass of the holes,\cite{ScBa10,RiWi08} when
compared to the electrons, and their associated tendency to be
localised at In-N-In chains, as shown by DFT
calculations.~\cite{LeWa2006,LiLu2010} It is important to note that
the observed hole localisation features both in-plane as well as
along the $c$-axis are vastly different from a standard
continuum-based description.
When looking at configuration 13, the hole wave function localises near the
bottom QW/barrier interface. This situation might be expected from a
continuum-based description. However, a fully continuum-based
description would not account for the clearly visible in-plane
localisation effect, since in such an approach InGaN/GaN QWs are
usually treated as one dimensional systems. For configurations 9 and 20, the
wave functions are localised in regions clearly above the bottom QW
interface. This would also not be expected from a continuum
description. This strong localisation experienced by the holes
validates the aforementioned conclusion that random alloy
fluctuations significantly impact the system properties for as
little as 10\% indium in the QW.
The results shown in Fig.~\ref{fig:10_percent_WFs} also indicate that the
wave function overlap between electron and hole ground states is not
only affected by the spatial separation along the growth direction
but also by the spatial separation in the $c$-plane. We will come back to this
observation in Sec.~\ref{sec:Results_Experiment} where we discuss the
observed results with respect to experimental data.

Before turning to the 15\% indium case, we focus on the 25\% indium
results, shown in Fig.~\ref{fig:25_percent_WFs}, in the next step.
When looking at the electron charge densities (red) of the here
displayed configurations 4, 13 and 15, it is evident that the
electron wave functions are all localised by the well-width
fluctuation. This results from the increased built-in field in the
25\% indium case when compared with the 10\% indium case [cf.
Fig.~\ref{fig:10_percent_WFs}]. Since the well-width
fluctuation introduces an extra in-plane/lateral confinement for the
electron wave functions, one could expect larger variations in the
corresponding electron ground state energies when compared with the
10\% indium system, where the considered well-width fluctuations are
only of secondary importance. This is because different microscopic
configurations of the indium atoms in the well-width fluctuation
will lead to different effective confining regions for the
electrons. For instance, a concentration of indium in the centre of
the well width fluctuation (configuration 13) can lead to a ground
state with a very different energy from that of a configuration
where the indium is concentrated near the barrier material (configuration
15); the state near the barrier is effectively confined in a smaller
region and will have its energy increased by the effects of the
barrier. This picture of small changes in indium content leading to
large changes in confinement and energy is consistent with the data
shown in Fig.~\ref{fig:electron_hole_vary} (a), and discussed
above. Turning to the hole ground states, we find a similar
behaviour as in the 10\% indium case with strong localisation
features for each configuration. In configurations 13 and 15 the
hole wave function is localised close to the bottom QW interface,
while the hole ground state wave function in configuration 4 is
localised two monolayers above the lower QW interface. Due to the
increased built-in field in the 25\% indium case one would expect
that the hole wave functions are localised near the bottom
interface. In this sense, with the hole wave function not being
localised near the bottom QW interface, one could expect that
configuration 4 represents an extreme case. This is confirmed in the
large deviation of its ground state energy from the average ground
state energy, as displayed in Fig.~\ref{fig:electron_hole_vary}~(b).
When looking at the ``Top View'' of the electron and hole ground
state charge densities, we find that electron and hole wave
functions are separately localised due to the built-in field, random
alloy fluctuations and well width fluctuations. Again, the wave
function overlap between electrons and holes is not only affected by
the spatial separation along the growth direction but also in the
$c$-plane. When looking, for instance, at the charge densities of
the electron and hole wave functions from configuration 15, we find
that electron and hole wave functions are localised at similar
in-plane positions. This is in contrast to configuration 4 and 13,
where we are left with a clear spatial separation also \textit{in}
the $c$-plane. Again, we will come back to the importance of these
properties in Sec.~\ref{sec:Results_Experiment}.

We now turn to the 15\% indium case. Here we have selected
configurations 3, 4 and 8. The corresponding electron and hole
ground state charge densities are displayed in
Fig.~\ref{fig:15_percent_WFs}. As discussed above, the variation in
the hole ground state energies [cf.
Fig.~\ref{fig:electron_hole_vary} (b)] is comparable to the
variations observed in the 10\% and 25\% indium case, respectively.
The isosurfaces of the hole charge densities (green) displayed in
Fig.~\ref{fig:15_percent_WFs}, show also a similar behaviour as in
the 10\% and 25\% indium systems. However, in comparison with the
10\% or 25\% indium case, the electron ground state energies in the
15\% indium system show much larger variations [cf.
Fig.~\ref{fig:electron_hole_vary} (a)]. When looking at the
isosurfaces of the electron charge densities (red) for
configurations 3, 4 and 8, the origin of this behaviour becomes
clear. In terms of the importance of the well-width fluctuation, the
15\% indium case represents an intermediate situation. For example,
in the case of configuration 8, the electron wave function is mainly
localised inside the well width fluctuation, while configuration 4
shows still significant charge density contributions outside the
well-width fluctuation. From this one could expect that the energies
of these different configurations are very different, and indeed
this is confirmed by Fig~\ref{fig:electron_hole_vary} (a). In
summary, the presence of the well-width fluctuation in combination
with the built-in field explains the initially surprising result of
the stronger variation in the electron ground state energies for the
15\% indium case in our calculations.

\begin{figure*}[t!]
\centering\includegraphics{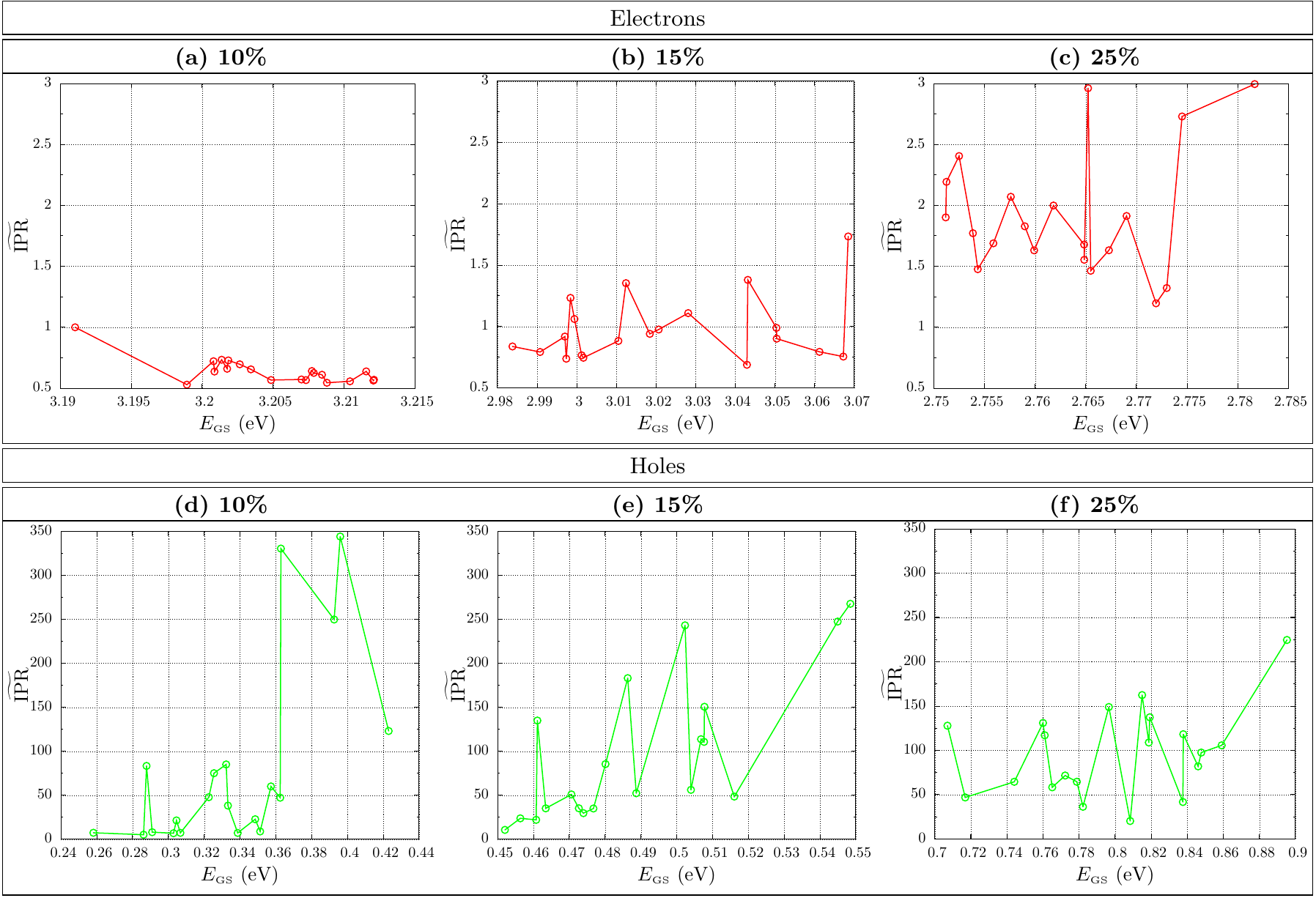}
\caption{Electron and hole ground state normalised inverse
participation ratios ($\widetilde{\text{IPR}}$) plotted as a
function of the ground state energies of each of the 20
microscopically different configurations, for nominal indium
contents of 10\%, 15\% and 25\%. The IPRs are normalised to that of
the 10\% electron ground state with the higest IPR, which is Config
13 (cf. Fig.~\ref{fig:10_percent_WFs}).} \label{fig:GS_PrVsE}
\end{figure*}
Overall, even though we have considered here only one particular
type of well-width fluctuation, our results clearly demonstrate that
their presence can contribute significantly to variations in both
transition energies and localisation effects. It should also be
mentioned that our results for electrons are consistent with the
work by Watson-Parris~\emph{et al.},~\cite{WaGo2011,ParrisThesis}
who studied the impact of well-width fluctuations on the electronic
structure of InGaN QWs in the framework of a modified effective mass
approach. In the study by Watson-Parris~\emph{et
al.},~\cite{WaGo2011,ParrisThesis} disk-shaped well-width
fluctuations with diameters ranging from 5~nm to 20~nm have been
studied. The influence of these fluctuations on the electron wave
function localisation characteristics has been analysed for
$c$-plane InGaN/GaN QWs with indium contents between 5\% and 25\%.
At 10\% indium content, the results with and without well-width
fluctuations are similar in terms of the electron ground state
localisation length. Only a slight decrease in the localisation
length is observed when the well-width fluctuations are included,
indicating that well-width fluctuations for lower indium contents
are of secondary importance, consistently with our results [cf.
Fig.~\ref{fig:10_percent_WFs}]. Watson-Parris~\emph{et
al.}~\cite{WaGo2011,ParrisThesis} showed also that at 25\% indium,
well width fluctuations lead to a significant reduction of the
electron ground state localisation length, when compared to a
calculation without well width fluctuations. This corroborates our
earlier mentioned conclusion that the importance of well width
fluctuations in localising the electron wave functions will depend
on the indium content. Therefore, even though we have assumed only
one particular type of well-width fluctuation, our presented results
provide a first indication of the importance of well-width
fluctuations on the electronic structure of $c$-plane InGaN/GaN QWs
with different indium contents.

%

So far our discussion of localisation effects has been based on
inspecting the charge densities of the electron and hole ground
state wave functions. To study localisation effects now on a more
quantitative basis we use the metric of the inverse participation
ratio (IPR).~\cite{Th1974} This provides a more objective measure of
localisation and also allows the examination and comparison of the
localisation characteristics of many states at once.
The participation ratio was first introduced by Bell~\cite{BeDe1970}
to assess the localisation properties of atomic vibrations. In that
context it gave insights into the fraction of the total number of
atoms in the system which participate effectively in the vibrations
of a particular normal mode. The IPR is the inverse of this
quantity, and is commonly used as a measure of localisation in TB
models.~\cite{Th1974,MeAh1986} In our TB formalism, a carrier wave
function, $\psi$, is given by:
\begin{equation}
\psi=\sum_i^{N}\sum_{\alpha}^{N_\alpha}a_{i\alpha}\phi_{i\alpha}\,\,
, \label{eq:WF_def}
\end{equation}
where the index $i$ runs over the $N$ lattice sites, and the index
$\alpha$ denotes the different orbitals in our $sp^{3}$ TB basis at
each site. The term $a_{i\alpha}$ represents the amplitude of the
wavefuntion, $\psi$, constructed with the basis
$\phi_{i\alpha}$, at the site $i$.
On the basis of
Eq.~(\ref{eq:WF_def}) the IPR may be defined as:
\begin{equation}
\text{IPR} =
\sum_{i=1}^{N}{\left(\sum_{\alpha}{\lvert a_{i\alpha}\rvert^{2}}\right)^{2}} /
\left(\sum_{i=1}^{N}{\sum_{\alpha}{\lvert a_{i\alpha}}\rvert^{2}}\right)^{2}\,\,
.\label{eq:TB_IPR}
\end{equation}
For a \emph{completely localised} state, which will be expressible
in terms of orbitals at only one atomic site, the IPR will be one;
for a \emph{completely delocalised} state, which is comprised of a
linear combination of equal parts of orbitals at all atomic sites,
the IPR will be $N^{-1}$; and for a state which is intermediate
between localised and delocalised, the IPR varies continuously
between one and $N^{-1}$.

In the following we have normalised the calculated IPR values to the
IPR value of the electron ground state with the largest IPR value
($1.529$x$10^{-4}$) in the 10\% indium case, which is configuration
13, shown in Fig.~\ref{fig:10_percent_WFs} (b). Therefore, the
normalised IPR values, $\widetilde{\text{IPR}}$, can exceed values
of one and can be interpreted as giving the extent to which the
state under consideration is more or less localised than the
electron ground state of configuration 13 shown in
Fig.~\ref{fig:10_percent_WFs}. Normalising the IPR values in
this way gives a more intuitive and visual picture of the localisation properties
of the state in question.

The ground state electron and hole $\widetilde{\text{IPR}}$ values
are shown as a function of their respective energies in
Fig.~\ref{fig:GS_PrVsE}. Figure~\ref{fig:GS_PrVsE} (a), (b) and (c)
correspond to the electron ground states in the 10\%, 15\% and 25\%
indium content systems, respectively. The data for the holes is
depicted in Fig.~\ref{fig:GS_PrVsE} (d), (e) and (f).
Figure \ref{fig:GS_PrVsE} confirms that the hole states are,
in general, far more localised than the electron states [\emph{please note
the different scales}]. More specifically, we find hole states which
are up to 350 times more localised than the electron ground state to which
they are normalised, and never less than 5 times more localised.
Furthermore, Fig.~\ref{fig:GS_PrVsE} reveals that the
$\widetilde{\text{IPR}}$ values of the hole states significantly
vary between different configurations, highlighting again that the
hole ground states are very sensitive to the microscopic alloy
structure.

To discuss the results in more detail, we start with the
$\widetilde{\text{IPR}}$ values of the electron ground states [Fig.~\ref{fig:GS_PrVsE} (a), (b), and (c)]. When comparing the electron
$\widetilde{\text{IPR}}$ values for the different indium contents,
we find that in general the $\widetilde{\text{IPR}}$ values increase
with increasing indium content. Since the macroscopic strain increases with increasing indium content, the piezoelectric built-in field
increases as well. Thus, with increasing indium content an
increasing confinement (along the $c$-axis)  for the
electron ground state can be expected. Additionally, with increasing indium content, the considered
well-width fluctuation becomes increasingly important and adds an extra in-plane
confinement. This is consistent with the trends observed across Fig.~\ref{fig:GS_PrVsE}(a-c) for the electron $\widetilde{\text{IPR}}$ values.
\begin{figure*}[t!]
\centering\includegraphics{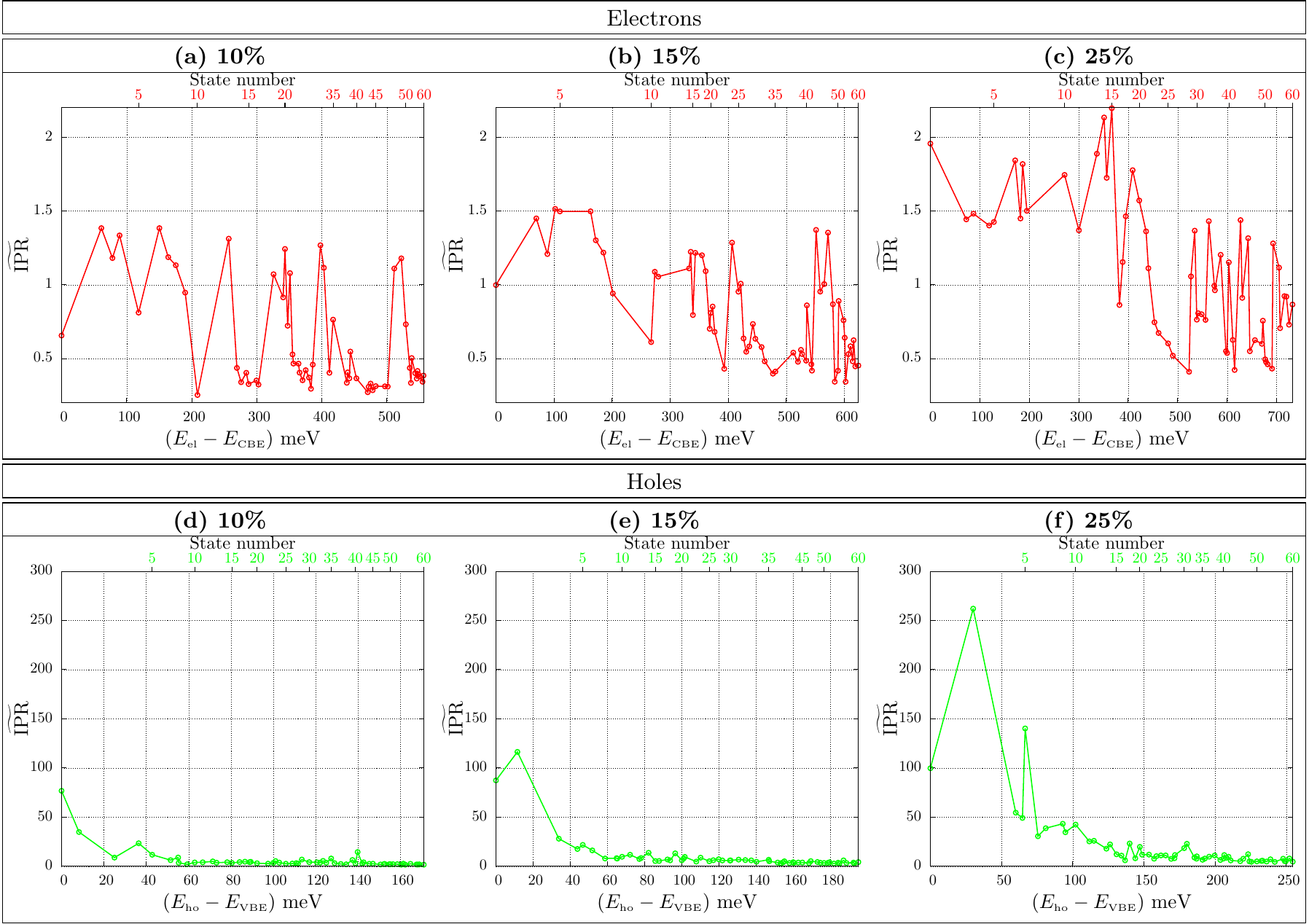}
\caption{Normalised electron (top row) and hole (bottom row) inverse participation ratios ($\widetilde{\text{IPR}}$)
plotted against the state energy as measured from the conduction or valence band edge. The results are given for particular
representative configurations with indium contents of 10\%, 15\% and 25\% (see text for selection criteria). The IPRs are normalised with respect to the IPR of the
most localised 10\% electron ground state, which is configuration 13.}
\label{fig:PrVsExcited}
\end{figure*}
However, when looking at the holes [Fig.~\ref{fig:GS_PrVsE} (d), (e), and (f)], we see that this trend is not as clearly
visible as in the electron case. This arises from several factors.
For instance, in the 10\% indium case [cf. Fig.~\ref{fig:GS_PrVsE} (d)], we have three exceptionally strongly
localised states. Their $\widetilde{\text{IPR}}$ values are of the
order of, or even exceed, the maximum values of the 15\% [Fig.~\ref{fig:GS_PrVsE} (e)] and 25\% [Fig.~\ref{fig:GS_PrVsE} (f)]
indium case. Therefore, to treat these exceptional states
accurately, a larger number of configurations would have to be
considered to perform more reliable statistical averages.
However, this is beyond the scope of the present study. Here we are
mainly interested in identifying general trends and to gain first
insights into the effects of random alloy fluctuations on the
electronic structure of $c$-plane InGaN/GaN QWs with varying indium
contents.

Based on the results presented, the argument of an increased built-in field with
increased indium content, used to explain the trends in the electron
ground states, cannot be directly applied to the hole states. The
reason for this is that the hole states show not only a strong
localisation along the growth direction, it is also evident from
Figs.~\ref{fig:10_percent_WFs},~\ref{fig:25_percent_WFs}
and~\ref{fig:15_percent_WFs} that the hole localisation has a very
strong in-plane localisation component. This component is not
greatly affected by the presence of the \emph{macroscopic} built-in
field along the growth direction. Thus the localisation behaviour of
the hole states is less dominated by the macroscopic built-in field
and governed more by fluctuations in the local indium environment. Consistent with
this, we find very large changes in the $\widetilde{\text{IPR}}$ values
of the ground state hole wave functions between
different configurations at nominally the same indium content, even
though the macroscopic built-in field should be very similar for a fixed indium content. Further to this, we note a tendency for localisation ($\widetilde{\text{IPR}}$) to increase
as the holes become more strongly confined, with a general rise from left to right in each
of Fig.~\ref{fig:GS_PrVsE} (d),(e), and (f); nevertheless, the fluctuations in the $\widetilde{\text{IPR}}$ values
are about as large as the trend itself.

So far we have focused our attention on localisation effects in
ground state properties. This provides key information for
experiments performed at low temperatures. However, when the optical
properties of $c$-plane InGaN/GaN QWs are studied experimentally at
ambient temperature, or when InGaN based devices are operating at room temperature, excited states become relevant. Thus a knowledge of the localisation characteristics of excited electron and hole states
is also important for understanding $c$-plane InGaN QWs. This is the focus of the
next section.

\subsection{Excited states}
\label{sec:ES_prop}

After studying the ground state localisation properties by means of
the $\widetilde{\text{IPR}}$ values, here we apply the same metric to the excited
states. We start by investigating, in Fig.~\ref{fig:PrVsExcited}, selected configurations for 10\%,
15\% and 25\% indium before looking at results averaged over the 20
different random configurations, considered for each composition. The benefit of studying selected
configurations first is that we can then display the results both as
a function of the energy and the state number. This is not possible
for the averaged data where the data is best displayed as a
function of the state number. This stems from the fact
that the ground state energies fluctuate significantly
between different configurations [cf. Fig.~\ref{fig:electron_hole_vary},\ref{fig:GS_PrVsE}].
The configurations selected here have a ground state
$\widetilde{\text{IPR}}$ value which is close to the average
$\widetilde{\text{IPR}}$ of all the ground states of that indium content, for electrons and holes.
Figure~\ref{fig:PrVsExcited} shows the $\widetilde{\text{IPR}}$
values for the first 60 electron and hole states as a function of
the energy, measured with respect to the corresponding ground state
energy. We take the absolute value of this energy difference so that
with increasing energy the states move deeper into the valence or
conduction band. The state numbers are given on the second $x$-axis
at the top of each figure.
Figure~\ref{fig:PrVsExcited} reveals a greater energy range covered
by the first 60 electron states than by the 60 hole states. This is due
to the larger hole effective mass in comparison
with the electron effective mass. This results in a smaller spacing between two
sequential hole states than for electrons and consequently 60 hole
states cover a smaller energy range than 60 electron states.

Looking at the electron states first, we note that in general the
$\widetilde{\text{IPR}}$ values for the first few excited states increase
with increasing indium content [Fig.~\ref{fig:PrVsExcited}(a), (b) and (c)]. We attribute this effect
to the increasing piezoelectric built-in field with increasing
indium content.
\begin{figure*}[t!]
\centering\includegraphics{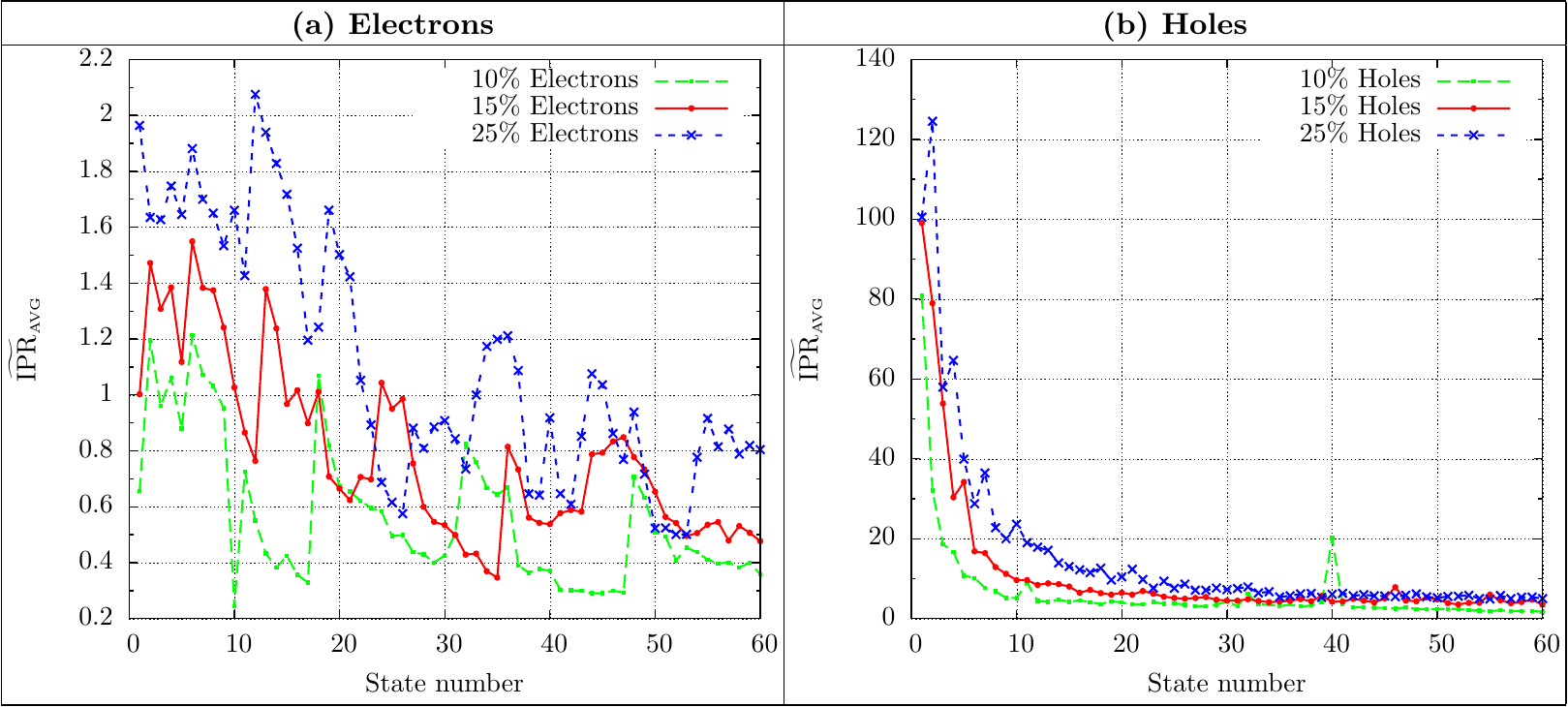}
\caption{Normalised inverse participation ratios of increasing electron (a) and hole (b) states averaged over
all microscopic configurations, $\widetilde{\text{IPR}}_{\text{\tiny{AVG}}}$, plotted against state number.
The results are shown for three different indium contents, 10\%, 15\%, and 25\%, and the normalisation
is taken with respect to the most localised 10\% electron ground state, which is that of configuration 13, shown in Fig.~\ref{fig:10_percent_WFs} (b).} \label{fig:AveragePr}
\end{figure*}
A similar trend is also observed for the hole states [Fig.~\ref{fig:PrVsExcited} (d), (e) and (f)]. For instance, the
$\widetilde{\text{IPR}}$ values of the first 5 hole states increase
with increasing indium content. Thus, one can expect
that for the holes, the energy depth into the valence band to which
there are still localised states found, increases with increasing
indium content. This is consitent with the experimentally observed increase of the PL width, stokes shift, and absorption edge broadening with indium content.~\cite{GrSo05,NaSa02,DoMa99,DoMa01,MaMi99}
The localised states in this energy range are sometimes referred to as ``tail states'' due to the
manner in which they modify the form of the density of states; the localised states appear as an added tail at
the begining of the ideally step-like 
form of the
density of states.~\cite{Ur1953,SrSa1986}
In order to gain some first insights into the relation
between localisation effects and energy, we note that for 10\%
indium [Fig.~\ref{fig:PrVsExcited} (d)] up to
$40$~meV into the valence band (state number 5) the holes have still
an $\widetilde{\text{IPR}}$ value which is approximately 10 times
larger than that of the highest $\widetilde{\text{IPR}}$
value of the electron ground state for the 10\% indium case.
Noting that this electron ground state shows considerable localisation [cf. Fig.~\ref{fig:10_percent_WFs}, Config 13],
we can safely say that a state 10 times more localised than this is strongly localised.
For 15\%
indium [Fig.~\ref{fig:PrVsExcited} (e)], again using
$\widetilde{\text{IPR}}\approx10$ as an arbitrary threshold,
this lasts up to $100$~meV (state number 20), and for 25\% indium
[Fig.~\ref{fig:PrVsExcited} (f)] there are still states with
$\widetilde{\text{IPR}}$ $\approx$ 10 past 165~meV at state 26. This behaviour, combined with the wide
variation in calculated ground state energy observed in Fig.~\ref{fig:electron_hole_vary}, implies that
a significant density of localised states
is present in $c$-plane InGaN/GaN QWs, which should measurably affect the optical properties
of these systems at elevated temperatures.

So far we have focused our attention only on selected
configurations.
To illustrate the generality of the observed behaviour in the excited
states, Fig.~\ref{fig:AveragePr} displays the
$\widetilde{\text{IPR}}$ values for the first 60 (a) electron and
(b) hole states averaged over the 20 different configurations.
These averaged normalised IPR values are denoted by $\widetilde{\text{IPR}}_{\text{AVG}}$.
The results are shown for 10\% (dashed green line), 15\% (red solid
line), and 25\% (dashed-crossed blue line) indium. Overall it is evident
that the hole states show much stronger localisation effects [higher
$\widetilde{\text{IPR}}$ values] when compared with the electrons.
Figure \ref{fig:AveragePr} also corroborates the earlier observed trend
that with increasing indium content there is an increased persistence of localisation
effects into the valence and conduction bands.

In an infinite system, the energy beyond which there are no more localised states can be expected at a definite energy, $E_{\text{mob}}$,
referred to as the mobility edge. However, even when using periodic boundary conditions, one is left
with a system of finite size. While each of these systems represents a portion of the real QW, there remain
finite-size effects that would not be present in the real, laterally infinite, QW.
For example, even though the highest valence state generally has a
very high IPR value in our finite-sized supercells, the most weakly bound of these states may be resonant
with delocalised QW states in the infinite system. We refer to such
states of our finite systems as ``quasi-localised'' states. These states should be excluded in
estimations of the energy range of localisation.
Taking this into consideration, we can
nevertheless combine the results obtained for the ground
and excited states in order to gain a first estimate of the
energy range of localised states which exist in the valence band
before the onset of delocalisation. We refer
first to Fig.~\ref{fig:GS_PrVsE} where it is evident that the ground state
energies for the holes vary by at least 100 meV across
different random indium confgurations.
We further note
from Fig.~\ref{fig:GS_PrVsE} that all of these hole states have very high
IPR values and can thus be considered strongly localised in comparison with the electrons.
Taking a conservative measure for the localisation depth
and keeping in mind the "quasi-localisation" effect described above, we consider only the
four energetically highest valence states for each
indium concentration studied to estimate the depth of localisation in the valence band.
This gives us approximately an energy spread of 50-60 meV.
\begin{figure*}[t!]
\includegraphics{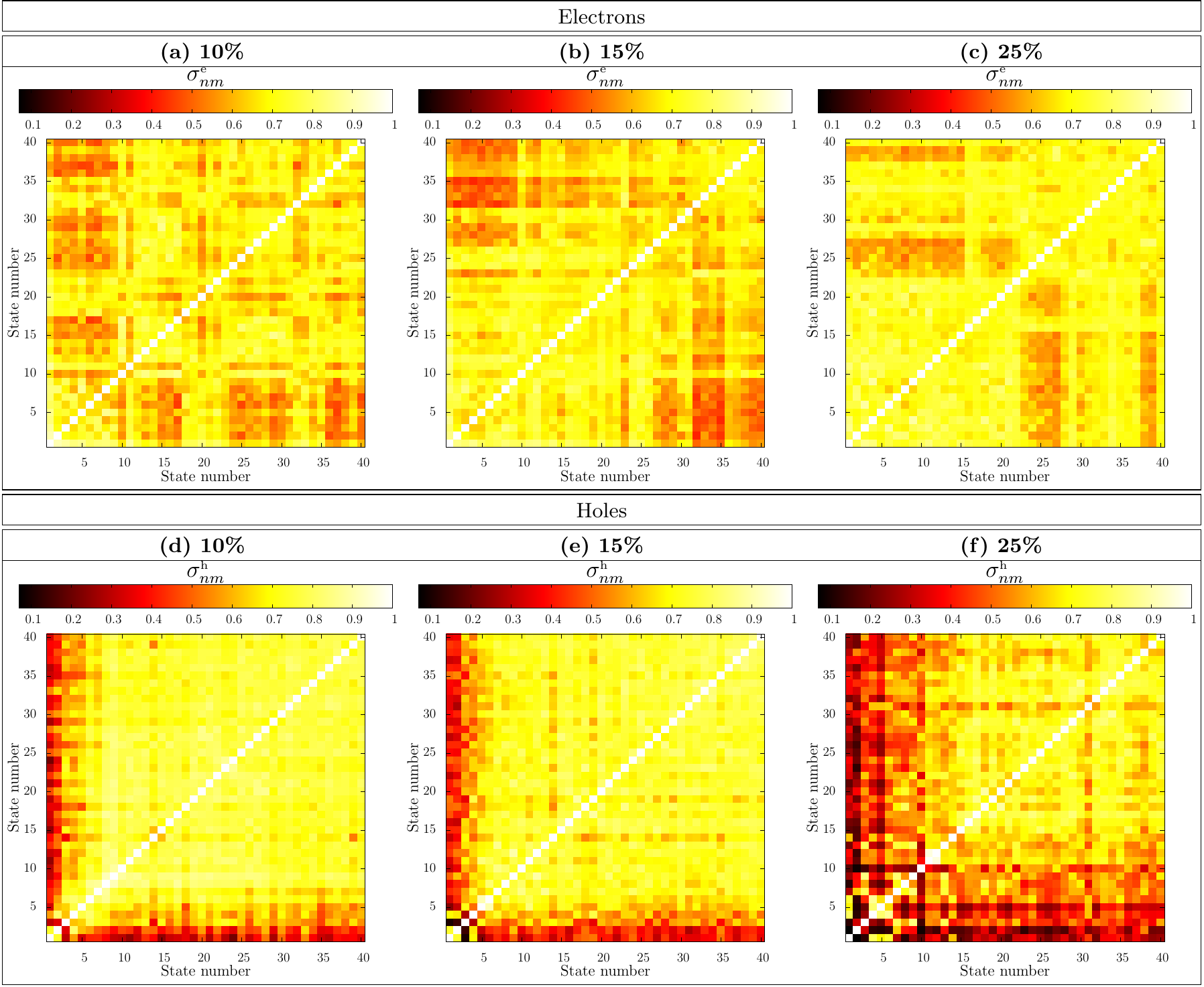}
\caption{Modulus wave function overlaps, $\sigma_{nm}^{\textrm{\tiny{e,h}}}$, of the first 40 states for electrons and holes
in particular configurations of In$_{x}$Ga$_{1-x}$N/GaN QWs with indium contents of 10\% ($x$=0.1), 15\% ($x$=0.15), and 25\% ($x$=0.25). The magnitude of the modulus overlap between state
$n$ and $m$ is indicated by the color of the point ($n$,$m$) on the plot.} \label{fig:SingleConfOverlaps}
\end{figure*}

Turning now
to the excited state data, we see, from the selected configurations studied in Fig. \ref{fig:PrVsExcited}, that, for instance,
in the 10\% indium case strong localisation effects ($\widetilde{\text{IPR}}_{\text{AVG}}>10$) extend
for at least an energy range of 40 to 50 meV below the
hole ground state. With increasing indium content this range
further extends [cf. Figs.~\ref{fig:PrVsExcited} (e) and (f)]. That these are not atypical
behaviours in our ensemble is supported by reference to
Fig.~\ref{fig:AveragePr}. Hence, combining the conservative estimate for the energy range of localised
states inferred from our ground state data and the insights from the excited state
studies, we estimate that already in case of 10\% indium a total spread of localised states amounts
to $\approx100$~meV. Thus, we expect
an energy range of at least 100 meV over which there
will be a significant density of localised valence states in
$c$-plane InGaN/GaN QWs with indium contents at above or equal to 10\%.

\begin{figure*}[t]
 \centering\includegraphics{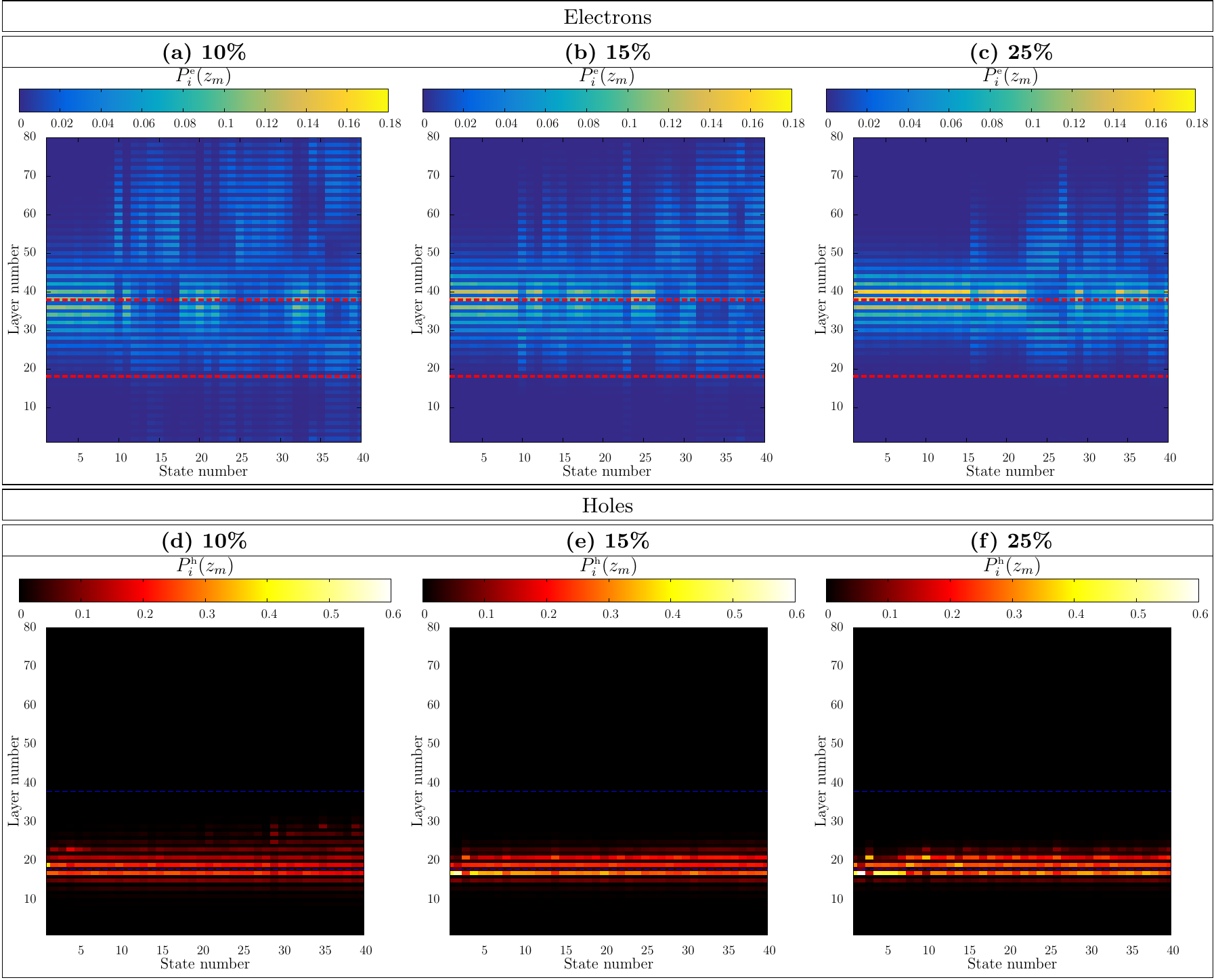}
 \caption{Planar integrated probability densities of electrons, $P_{i}^{\textrm{\tiny{e}}}(z_{m})$, [(a),(b) and (c)]
 and holes, $P_{i}^{\textrm{\tiny{h}}}(z_{m})$, [(d),(e) and (f)]
 in $c$-plane InGaN quantum wells with indium contents of 10\%, 15\% and 25\%, for the first 40 states.
 The index $i$ refers to the state number,
 and the index $m$ refers to the layer of the quantum well. The boundaries of the quantum well
 active region are indicated by blue dashed lines on the figure. To emphasise the different colour scales,
 a different colour scheme has been used for electrons and holes.
 } \label{fig:Layer_Loc}
 \end{figure*}
Even though we cannot determine the density of localised states
exactly, we can still analyse how the wave function overlap between
carriers in different states is affected by alloy fluctuations. This
question is, for example, relevant for transport through $c$-plane
InGaN/GaN QWs, since it gives a first indication of the probability
of a carrier transferring from one site/state to another. To gain
some insight into the carrier overlap we study in the following the
\emph{modulus overlap} of the wave functions of different states. In
our TB formalism the modulus overlap $\sigma_{nm}$ between two
states $\psi_{m}$ and $\psi_{n}$ can be defined by:
\begin{equation}
\sigma_{nm} =
\sum_{i}^{N}\lvert\psi_{n,i}\rvert\lvert\psi_{m,i}\rvert\,\, ,
\end{equation}
where $i$ denotes the lattice site. For $n=m$, this will be the
overlap of a state with itself, $\sigma_{nn}=1$, since the wave
functions are normalised. A state $\psi_{j}$ with a large
$\sigma_{jm}$ value for many states $\psi_{m}$ will then have a
widely spread out wave function. Conversely, if a state $\psi_{j}$
has a small $\sigma_{jm}$ for many other states, $\psi_{m}$, it
means that the wave function $\psi_{j}$ is localised in a particular
spatial region of the QW. Note that we are dealing here with the
\emph{modulus} overlap; our definition for the overlap does not take
into account the parity of the respective wave functions. Our metric
simply indicates the extent to which the densities of the involved
carriers are spatially coincident.

Figure~\ref{fig:SingleConfOverlaps} (a),(b), and (c) show, for the
same configurations chosen in Fig.~\ref{fig:PrVsExcited}, the
modulus overlaps, $\sigma^{e}_{nm}$, of each electron state with every
other electron state. The data for the hole states are displayed in
Fig.~\ref{fig:SingleConfOverlaps} (d),(e) and (f). We have
considered the first 40 electron and hole states. The left column, (a) and (d),
contains the results for 10\% indium, the middle column, (b) and (d), the data for
the 15\% indium case while the right column, (c) and (f), depicts the situation
for the QW with  25\% indium. 

We begin our analysis by focusing on the modulus hole wave function
overlap $\sigma^h_{nm}$ [Fig.~\ref{fig:SingleConfOverlaps} (d), (e)
and (f)]. In the 25\% case, (f), there are three distinct regimes in
$\sigma^h_{nm}$ visible. Over the first five states a dark region of
very poor overlap [small $\sigma^{h}_{nm}$ value] is visible. This
indicates a region of strongly localised states, with the hole
localisation length well below the supercell size (10~nm) considered
here. This is consistent with the very high $\widetilde{\text{IPR}}$
values shown in Fig.~\ref{fig:GS_PrVsE}. After the first five
states, from state 6 to 10, we find a region of ``semi-localised''
states, with $\sigma^h_{nm}$ values around 0.3 to 0.7. Beyond these
states we find an area in the $\sigma^h_{nm}$ plot that has values
between 0.7 and 1. We classify these states as ``delocalised
states''. Looking at the 15\% indium case
[Fig.~\ref{fig:SingleConfOverlaps} (e)] we find again these three
regions but with the ``delocalised'' region being much larger, and
both the ``semi-localised'' and ``localised'' region being greatly
reduced. For the 10\% indium case it is very hard to discern a
region which could be described as ``strongly localised'' in the
same sense as for 25\% indium, but there is still clearly visible a
``semi-localised'' region. Consistent with our discussion of
Figs.~\ref{fig:PrVsExcited} and~\ref{fig:AveragePr}, it can be
concluded that for the holes the location of the ``mobility edge''
depends on the indium content.

\begin{figure*}[t!]
 \includegraphics[width=\linewidth]{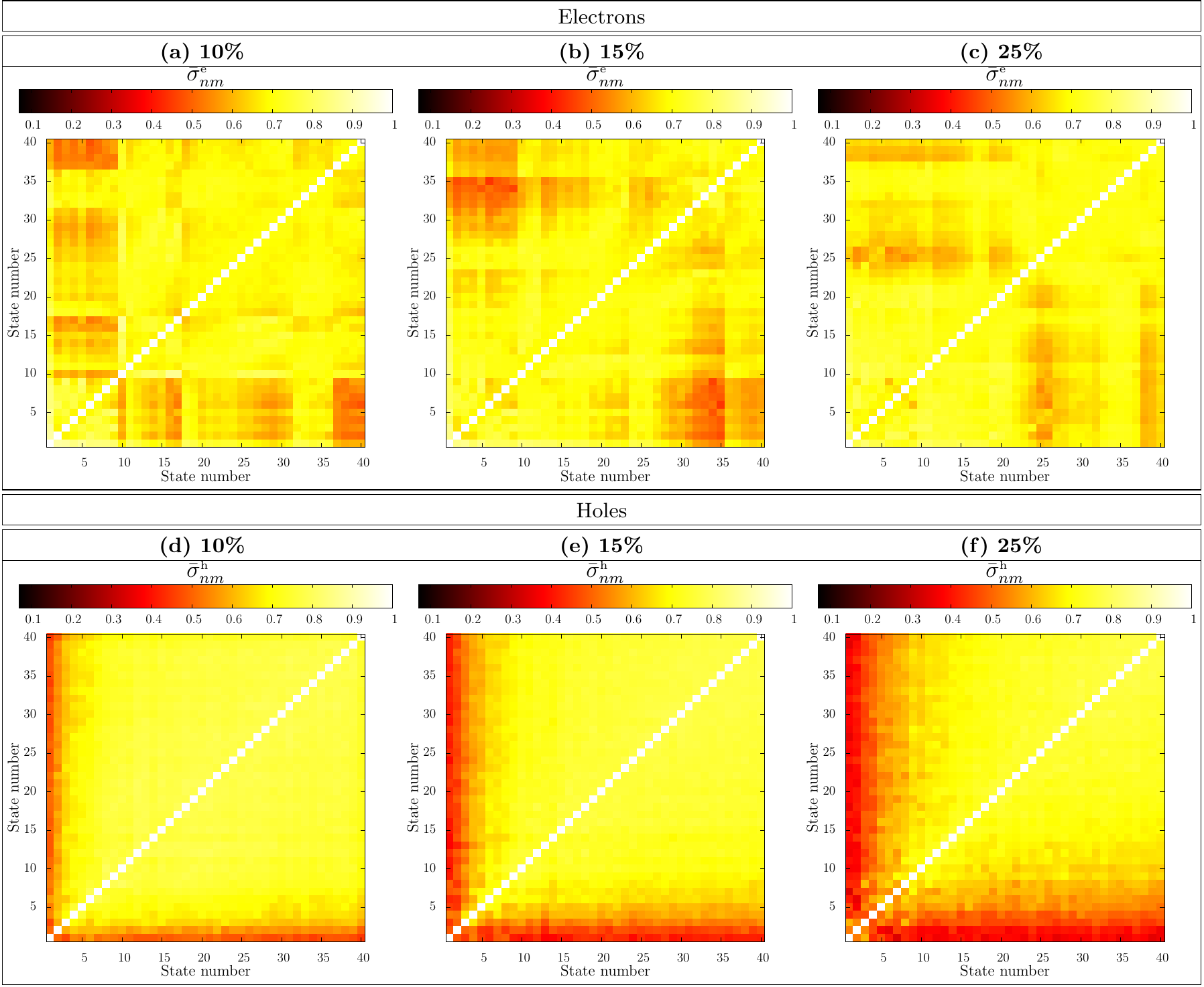}
 \caption{Modulus wave function overlaps of the 40 electron (top row) and hole (bottom row) states, of In$_{x}$Ga$_{1-x}$N/GaN
 quantum well systems averaged over 20 different microscopic configurations. The averaged overlaps are denoted by $\bar{\sigma}^{e,h}_{nm}$. The data are shown for three different indium contents, 10\%, 15\% and 25\%.}
 \label{fig:Overlaps}
\end{figure*}
Looking at the electron states, in the upper row of
Fig.~\ref{fig:SingleConfOverlaps}, the minimum values
of $\sigma^e_{nm}$ are much larger [$\sigma^{e}_{nm}\approx0.4$] when
compared to the holes [$\sigma^{h}_{nm}\approx0.05$]. We attribute this to
the fact that the electron states, as discussed in
Sec.~\ref{sec:GS_prop} and demonstrated in
Fig.~\ref{fig:PrVsExcited}, are less perturbed by the alloy
fluctuations. The light and dark
overlap ``bands'' of the figure correspond to the overlaps between
states localised inside the QW and states which start to spread into the GaN barrier material. When
looking at the positions (state numbers) of the dark regions with
low $\sigma^{e}_{nm}$ values, we find that these regions in general
move to higher state numbers with increasing indium content. The different positions (state numbers) of these ``bands''
for the different indium contents can be imputed to changes in the conduction band
confinement potential. Please note that with increasing
indium content, the electron wave functions also begin to become
localised by the well-width fluctuations, which then also affects $\sigma^{e}_{nm}$.

To support these arguments and to further clarify the features seen in the
$\sigma^{e}_{nm}$ values, we now present the planar integrated
probability density, $P_i$, of each state $\psi_i$:
\begin{equation}
P_i(z_m) = \sum_{k,l}|\psi_i(x_k,y_l,z_m)|^2\,\, ,
\end{equation}
where $x_k$ and $y_l$ are the in-plane ($c$-plane) indices and $z_m$
denotes the layer index along the $c$ axis.  The quantity $P_i(z_m)$
gives the probability that the electron or hole state $i$ be
found in the layer specified by the index $z_{m}$. This allows us to
shed light on the localisation characteristic of the different
states along the $c$ axis.
\begin{figure*}[t!]
 \includegraphics[width=\linewidth]{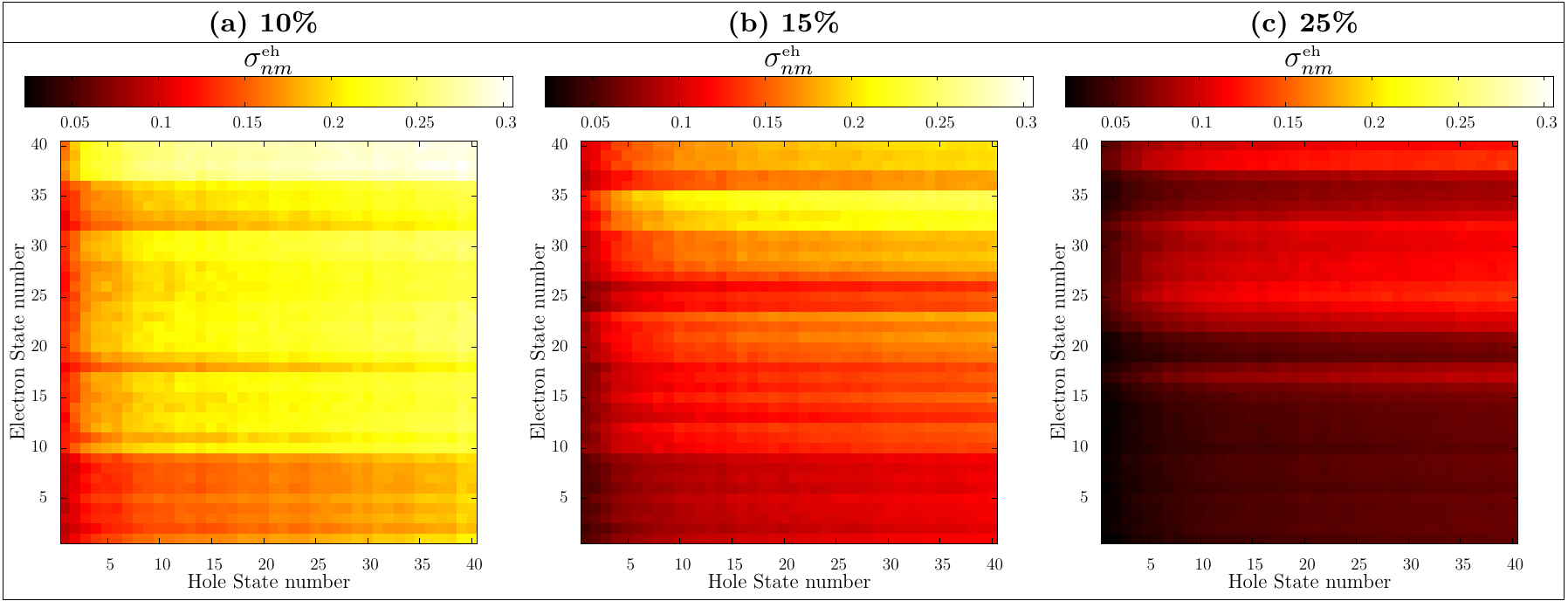}
 \caption{Modulus wave function overlaps of the 40 electron and hole states, in In$_{x}$Ga$_{1-x}$N/GaN
 quantum well systems averaged over 20 different microscopic configurations. The averaged overlaps are denoted by $\bar{\sigma}^{eh}_{nm}$. The data are shown for three different indium contents, 10\%, 15\% and 25\%.}
 \label{fig:ehOv}
\end{figure*}

Figure~\ref{fig:Layer_Loc} shows $P_i(z_m)$ plotted for the
first 40 electron [upper row] and hole states [lower row] for 10\%,
15\% and 25\% indium [left to right in Fig.~\ref{fig:Layer_Loc}]. In
each figure, the horizontal axis denotes the state number and $P_i(z_m)$ is
given on the vertical axis. Thus the point (1,2) will give the
probability that the electron/hole described by the first eigenstate be found in the
second layer of the supercell. As a guide to the eye we indicate the
QW boundaries as (blue) dashed lines. Figure~\ref{fig:Layer_Loc}
confirms that much of the structure observed in the $\sigma^{e}_{nm}$ values displayed
in Figs.~\ref{fig:SingleConfOverlaps} (a-c), arises from variations in overlap with increasing height, $z_{m}$, in the QW.
Conversely, Fig.~\ref{fig:Layer_Loc} indicates that for the holes it is primarily the in-plane separation and in-plane variation in overlap, rather than the separation
in $z_{m}$, that leads to the structure observed in $\sigma^{h}_{nm}$ shown in Figs.~\ref{fig:SingleConfOverlaps} (d-f).

So far we have focused our discussion on selected configurations to
illustrate trends in the localisation characteristics of ground and
excited states. In order to demonstrate the generality of these
results we have calculated the modulus wave function overlap
averaged over all configurations for each
of the indium contents considered here. This is denoted by $\bar{\sigma}^{e}_{nm}$ for electrons, and $\bar{\sigma}^{h}_{nm}$ for holes. The
$\bar{\sigma}^{e,h}_{nm}$ results are displayed in
Fig.~\ref{fig:Overlaps}, which shows that $\bar{\sigma}^{e,h}_{nm}$
reflects the trends observed in the selected configurations [cf.
Fig.~\ref{fig:SingleConfOverlaps}]. Notably, the overlap structure
in the electrons is preserved. For the holes the different regimes
of localisation are again apparent across the different indium
contents, especially the increasing width of the
``strong-localisation'' region with increasing indium content.

To gain further insight into the impact of random
alloy fluctuations and varying indium content on the electronic and
optical properties of $c$-plane InGaN/GaN QWs, we analyse in a next step the
modulus overlap, $\sigma^{eh}_{nm}$, of the first 40 electron and hole
wave functions. The results are shown in Fig.~\ref{fig:ehOv}. The
data give first indications of how the emission efficiency of the
QWs are affected by changes in the indium content. Our results reveal that
with increasing indium content the electron and hole modulus wave
function overlap decreases. This effect can be attributed primarily to the
increasing strain-dependent macroscopic piezoelectric polarization
field in the QW. Consequently, one observes a stronger spatial
separation of the electron and hole wave functions along the
$c$-axis. However, the increasing in-plane localisation introduced 
by well width and alloy fluctuations, which also increase with increasing indium content (from 10\% to 25\%),
as shown earlier, will also contribute to an increasing reduction in overlap. 
This finding is consistent with previous experimental studies on InGaN/GaN
QWs, wherein the internal quantum efficiency decreases rapidly with increasing indium content for long emmission wavelengths.~\cite{MuYa99} However, our data indicates also that when looking at
$\sigma^{eh}_{nm}$ for a fixed indium content, the modulus overlap
is state number dependent and increases in general with increasing
state number. For instance, when looking at the electron ground
state, $n=1$, in the in 10\% In case, we find a low $\sigma^{eh}_{1m}$
value for hole states with $m<5$, while the value is clearly larger
when $m>35$. This indicates a more efficient radiative recombination
rate for transitions involving the electron ground states and excited hole states. We
attribute this behavior to the effect observed in
Fig.~\ref{fig:Overlaps}, wherein regions of
``delocalised hole states'' are found.

\section{Comparison with experimental data}
\label{sec:Results_Experiment}

In a first step of our theory-experiment correlation
we analyse the calculated average ground state transition energies
and compare them with measured PL peak energies from the
literature.~\cite{GrSo05} The data are summarised in
Table~\ref{tbl}. The here considered indium contents $x$ are very close
to the experimental values; however, we have kept the well width $L$
constant, while in the experiment this quantity varies between the
different samples. Nevertheless, the reported theoretical data
is in good agreement with the experimental values, and given that our
well width in general is larger than the experimental value, we
slightly underestimate the PL peak energies.
\begin{table}[t]
\small
  \caption{\ Comparison between calculated average ground
state transition energies (Calc) and experimental PL peak energies (Exp) obtained at low temperatures ($T=6K$) in
In$_{x}$Ga$_{1-x}$N/GaN $c$-plane QWs.~\cite{GrSo05} The QW well
width is denoted by $L$ and the transition energies/PL peak position
energies are given by $E_{g}$. }
  \label{tbl}
  \begin{tabular*}{0.5\textwidth}{@{\extracolsep{\fill}}llll}
    \hline
     & $x$ & $L$~(nm) & $E_{g}$~(eV) \\
    \hline
     Exp & 0.25 & 3.3 & 2.162\\
     Calc & 0.25 & 3.5 &1.964 \\ \hline
     Exp & 0.15 & 2.9 & 2.707\\
     Calc & 0.15  & 3.5 &2.533 \\ \hline
     Exp & 0.12 & 2.7 & 2.994\\
     Calc & 0.10 & 3.5 &2.871\\ \hline
    \hline
  \end{tabular*}
\end{table}

Our theoretical findings on the nature of the
localisation in these systems also support several experimental
studies and their proposed explanations. For instance, the
experimentally observed shift in the PL peak position with
temperature, usually referred to as the ``S-shape'' dependence, is
normally attributed to the existence of localised carriers in
$c$-plane InGaN/GaN QWs. Based on our data we can conclude that
localisation effects play not only a significant role in ground but
also in excited hole states. This is even the case in $c$-plane
InGaN/GaN QWs with as little as 10\% indium. From our data we expect
an energy range of order $100$~meV over which there will be a
significant density of localised valence states in $c$-plane
InGaN/GaN QWs. This is consistent with the minimum energy ranges
which can be inferred from the blue shift due to thermal
redistribution amongst localised states in temperature dependent PL
and electroluminescence (EL) experiments, which ranges from
$55-200$~meV~\cite{ElPe97,ChGa98} for different QW structures.

In addition to the ``S-shape'' dependence of the PL and EL spectra,
time dependent PL spectra of $c$-plane InGaN/GaN QWs show
non-exponential decay transitions and the measured decay times vary
across the spectrum.~\cite{DaSc16,DaDa00} An explanation for this
behaviour has been put forward by Morel \emph{et al.},~\cite{MoLe02}
using a model of independently localised electron and hole wave
functions. Under this assumption, Morel \emph{et al.}~\cite{MoLe02}
were able to achieve very good agreement between theoretical
predictions and experimentally observed data. Our results support
the assumption of individually localised carriers as demonstrated in
Sec.~\ref{sec:GS_prop}. We show here that this behaviour is
independent of the considered indium content, since even at 10\%
indium both electron and hole charge densities show indications of
localisation effects. We have also shown in Ref.~\citenum{ScCa2015}
that localisation effects and built-in fields dominate over Coulomb
effects and thus a single-particle picture should already provide a
reliable description of the localisation features in $c$-plane InGaN
systems. It is important to note that the spatial separation between
electron and hole wave functions is not only affected by the
presence of the macroscopic built-in field but also by the
localisation characteristics of the wave functions in the $c$-plane.
Thus, while in a continuum based $c$-plane QW description, dipole
matrix elements are determined by the spatial separation along the
growth direction only, in our atomistic calculation the relative
position of the electron and hole wave functions within the
$c$-plane also plays an important role. As the relative in-plane
positions of electrons and holes change as a function of the
configuration [cf. Figs.~\ref{fig:10_percent_WFs}
-~\ref{fig:15_percent_WFs}] the dipole matrix elements will change
between different configurations, as demonstrated in Ref.
\citenum{ScCa2015}. This behaviour is consistent with the
non-exponential PL decay curves and the variation of decay time
across the spectrum.

\section{Conclusion}
\label{sec:Summary}

In summary, we have presented a detailed analysis of the electronic
structure of $c$-plane In$_x$Ga$_{1-x}$N/GaN QWs with indium
contents of $x=0.1$, 0.15 and 0.25,
covering the experimentally relevant range. To perform
this analysis we have used a fully atomistic description, including
local alloy, strain and built-in field variations arising from
random alloy fluctuations. In addition to going beyond the usually
applied continuum-based description for these systems, we give
insight into not only ground state properties but also excited state
properties.

From our analysis we conclude that for as little as 10\% indium in the QW,
the valence band structure is strongly affected by localisation
effects. Our results indicate that well width fluctuations could
lead to electron wave function localisation effects in addition to
localisation effects introduced by random alloy fluctuations. These
observations hold not only for ground states but also for excited
states. From an initial estimate of our data, we conclude that even
at 10\% indium in the well, we are left with an energy range of order
$100$~meV into the valence band that should be dominated by
strongly localised states. Our data also indicate that this energy range
increases with increasing indium content. Experimental
data, such as the ``S-shape'' dependence of the PL peak position
with temperature gives clear experimental evidence of the presence
of such (excited) localised states. Our theoretical findings are
therefore consistent with experimental observations.

Moreover, by looking at (modulus) wave function overlaps between the
first 40 hole or electron states, we gained initial insights into
the probability of transferring carriers from one site/state to
another. Our investigations indicated different regimes ranging from
strongly ``localised states'' up to ``delocalised states''. While
the localised states have very little overlap with all other states,
the delocalised states reveal a high overlap with most of the other
considered states. These features are relevant for experimental
studies at ambient temperature and transport properties. In
particular, the strong hole wave function localisation should affect
the hole transport in $c$-plane InGaN-based multi-QW LEDs
significantly. The observed localisation effects will impact both
the vertical transport along the $c$-axis through the different QWs,
and also the lateral transport and thus how the carriers spread
within the growth plane of the QW. Thus, these localisation features
are relevant in general for InGaN-based devices operating at room
temperature and above. The obtained data will now form the basis for
more detailed transport and in general device-related calculations.

Finally, our theoretical study showed that built-in field, random
alloy and well width fluctuations lead to the situation of
independently localised electron and hole wave functions in
$c$-plane InGan/GaN QWs. This holds for as little as 10\% indium in
the QW. This finding is consistent with the ``pseudo 2-D
donor-acceptor pair'' model proposed by Morel \emph{et
al.}~\cite{MoLe02} to explain time resolved PL measurements of
$c$-plane InGaN/GaN QWs.

\section*{Acknowledgments}
This work was supported by Science Foundation Ireland (project
numbers 10/IN.1/I2994 and 13/SIRG/2210) and the European Union 7th
Framework Programme DEEPEN (grant agreement no.: 604416). The
authors would like to thank Phil Dawson, Colin J.~Humphreys, Rachel
A.~Oliver, Alvaro Gomez-Iglesias, and Lutz Geelhaar for fruitful
discussions. Furthermore, computing resources provided by Science
Foundation Ireland (SFI) to the Tyndall National Institute and by
the SFI and Higher Education Authority funded Irish Centre for High
End Computing (ICHEC) are acknowledged.

\bibliographystyle{apsrev4-1}
\bibliography{./Daniel_Tanner_Bibliography}

\end{document}